\documentclass[reprint,amsmath,amssymb,aps,]{revtex4-1}
\usepackage{graphicx}
\usepackage{amsmath}
\usepackage{subfigure}
\usepackage{mathtools} 
\usepackage{color}
\usepackage{overpic}
\usepackage[export]{adjustbox}
\usepackage{mwe, tikz}
\usepackage{rotating}
\usepackage{latexsym}
\usepackage{csquotes}
\usepackage{tikz} 
\usetikzlibrary{calc,arrows,decorations.markings} 
\usepackage{pict2e}
\usepackage{pgfplots}
\usepackage{float}
\usepackage{hyperref}
\usepackage{xr}
\begin{document}
\title{From plastic flow to brittle fracture: role of microscopic friction in amorphous solids}
\author{Kamran Karimi}
\affiliation{Universit\'e Grenoble Alpes, CNRS, ISTerre, 38041 Grenoble cedex 9, France}
\author{David Amitrano}
\affiliation{Universit\'e Grenoble Alpes, CNRS, ISTerre, 38041 Grenoble cedex 9, France}
\author{J{\'e}r{\^o}me Weiss}
\affiliation{Universit\'e Grenoble Alpes, CNRS, ISTerre, 38041 Grenoble cedex 9, France}
\begin{abstract}
Plasticity in soft amorphous materials typically involves collective deformation patterns that emerge upon intense shearing.
The microscopic basis of amorphous plasticity has been commonly established through the notion of ``Eshelby''-type events, localized abrupt rearrangements that induce flow in the surrounding material via \emph{non}-local \emph{elastic}-type interactions.
This universal mechanism in flowing disordered solids has been proposed despite their diversity in terms of scales, microscopic constituents, or interactions.
However, we argue that the presence of frictional interactions in granular solids alters the dynamics of flow by nucleating micro shear \emph{cracks} that continually coalesce to build up system-spanning \emph{fracture}-like formations on approach to failure.
The plastic-to-brittle failure transition is uniquely controlled by the degree of frictional resistance which is in essence similar to the role of heterogeneities that separate the abrupt and smooth yielding regimes in glassy structures.       
\end{abstract}
\maketitle
%
\emph{Introduction-} Amorphous solids exhibit collective features upon application of stress.
A dense granular matter, for instance, forms a complex force network under compression with chain-like structures that are extended far beyond the grain size \cite{majmudar2005contact,karimi2011local}.
Upon shear failure, system-spanning strain features emerge which are commonly referred to as the ``shear-banding" phenomenon \cite{desrues2002shear}.
Emergence of the macroscopic failure is not abrupt but is commonly preceded by critical fluctuating patterns that have been recently described in the context of the so-called \emph{yielding transition} \cite{LinPNAS2014}.
The microscopic basis of this viewpoint is the appearance of recurring plastic bursts that are quite localized in space but have long-range elastic-type consequences \cite{picard2004elastic,le2014emergence}. 
In this framework, local isolated events are initially activated by external deformation (in the absence of thermal fluctuations), but then further instability may be triggered and propagates due to non-local interactions.
This \emph{avalanche}-like dynamics leads to the progressive formation of scale-free clusters near the failure transition which may be viewed as a continuous second-order phenomenon with unique scaling properties associated with it \cite{zapperi1997plasticity, karimi2017inertia}.

Due to structural heterogeneities and initial disorder, the failure patterns in glassy systems tend to be rather \emph{diffuse} with cascades of events that are highly intermittent and scattered both in time and space.
As the transition approaches, the rate of plastic activity accelerates and fluctuations become correlated covering larger and larger scales \cite{castellanos2018avalanche,gimbert2013crossover}.
In this manner, the yielding process involves a progressive coalescence of transient events that occur smoothly over an extended period of time prior to the main instability.   
On the other hand, in certain systems that lack the heterogeneity element, the nucleation process takes place sequentially creating spatially elongated formations that are marked with a spontaneous release of the accumulated elastic energy. 
In the latter case, no or very few detectable \emph{precursory} signals precede the main event, which bears undesired consequences in terms of failure forecasting \cite{vasseur2015heterogeneity}.
The so-called ``diffuse"-to-``localized" transition was also reported experimentally in \emph{Coulombic} materials where the failure modes appear to be controlled by the degree of frictional resistance \cite{jaeger2009fundamentals}. 
This view was also validated within the context of a continuum damage model and Mohr-Coulomb plasticity that weakly frictional systems favor more scattered damage distributions \cite{amitrano1999diffuse}.
In a more recent study \cite{karimi2018correlation}, in the framework of a standard elasto-plastic model \cite{nicolas2017deformation}, the permanent localization phenomenon in granular flow was rationalized via an emergent weakening mechanism that was attributed to a compromise between elastic-type couplings among small-scale events and the pressure-sensitivity in local yielding thresholds.
Recent granular experiments have analyzed the anisotropic structure of the mesoscopic patterns within the context of linear elasticity and the concept of shear transformations accompanied with local dilatancy \cite{le2014emergence}.  

\begin{figure}[b]
		\includegraphics[width=0.3\textwidth]{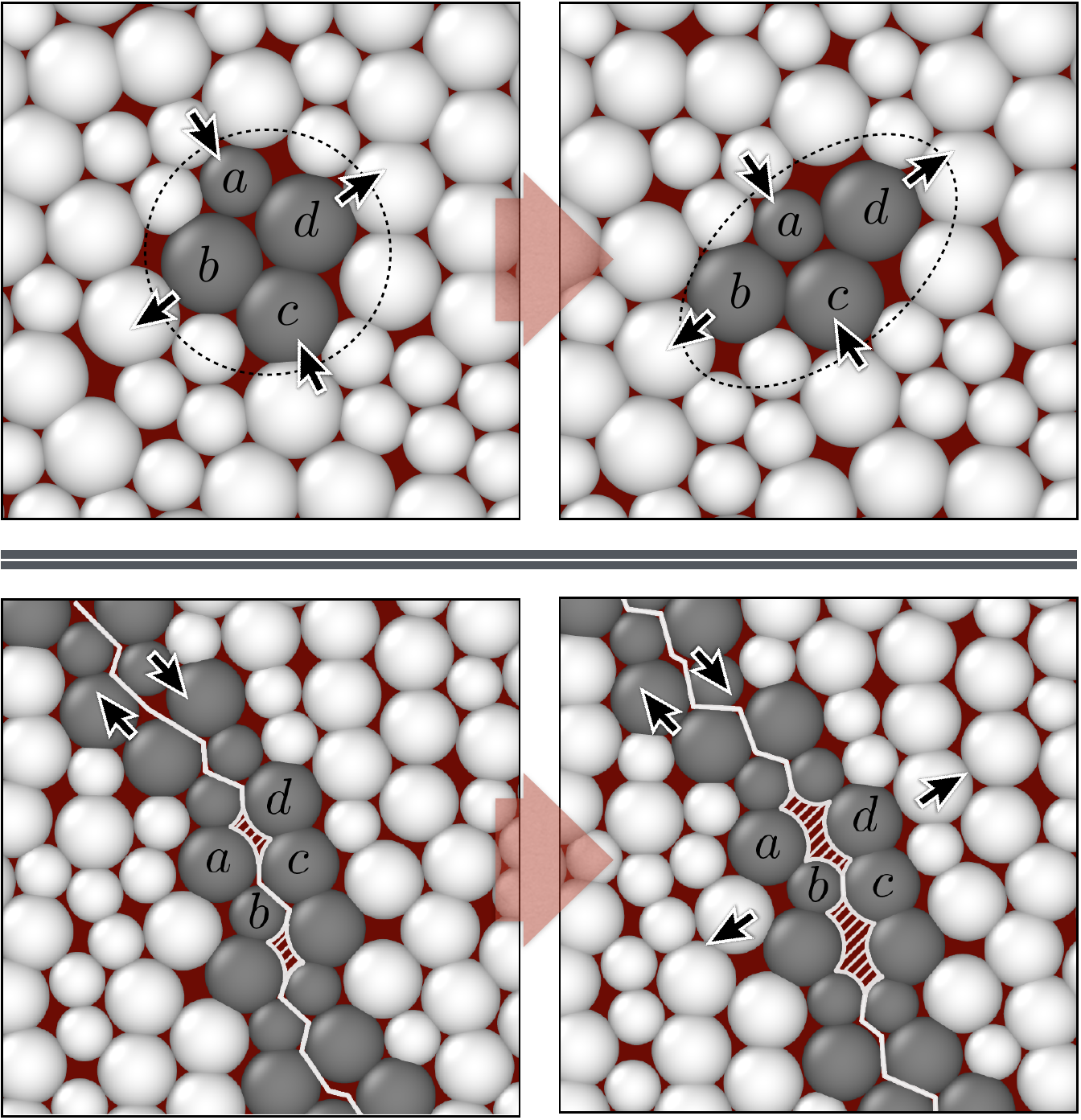}
	\caption{Slip kinematics at grain level. \textbf{Top}: Frictionless grains undergo localized rearrangements. The black arrows display the area-preserving nature of the transformation. \textbf{Bottom}: Frictional sliding typically involves a chain of grains that continually \emph{ride up} on one another producing significant ``dilatancy'' during slip. This effect is indicated by shaded areas in the sketch with solid lines marking the average shear plane.}
	\label{fig:flowDynamics}
\end{figure}

The aim of this paper is to make a distinction between two contrasting failure mechanisms based upon a purely microscopic point of view.
We base our analysis on the observation that the local dynamics of slip events is totally sensitive to the extent of frictional strength (see Fig.~\ref{fig:flowDynamics}).
In the absence of microscopic friction, flow is fully plastic with grains that continually rearrange and locally slip in a fashion quite similar to $``\text{T}1"$ process in sheared foams \cite{kabla2003local}.
Frictional flow, by contrast, is characterized by fast sliding of meso-scale blocks which go through strong dilation upon shearing analogous to ``Mode $\text{II}$'' fracture \cite{miannay1997fracture}.
The term dilatancy refers to the component of motion normal to the average slip plane that accompanies distortions as a result of the surface roughness. 
We verify the relevance of these elementary relaxation processes within our model by quantifying the structure of correlations that incur upon macroscopic failure.
We also argue that the grain-level friction governs the dynamics of avalanche-like precursors that are collective in nature and dictate the ``brittleness'' degree of the ultimate failure. 
This is analogous to the role of structural disorder in glass-like formations in that homogeneities potentially enhance the global stability of the initial configurations and, therefore, may alter the nature of transition {\cite{ozawa2018random, popovic2018elasto}.
Here, however, we span the transition not by changing the disorder strength but through local friction.
These findings may have important consequences in terms of the accuracy of the failure prediction. 

\emph{Bi-axial Compression Test}- We used bi-disperse packings of $N$ frictional disks in a bi-axial loading geometry illustrated in Fig.~\ref{fig:setUP} (see Supplementary Materials Sec.~\ref{sec:SimulationsAndProtocols}).
\begin{figure}
	\begin{center}
		\begin{overpic}[width=0.425\textwidth]{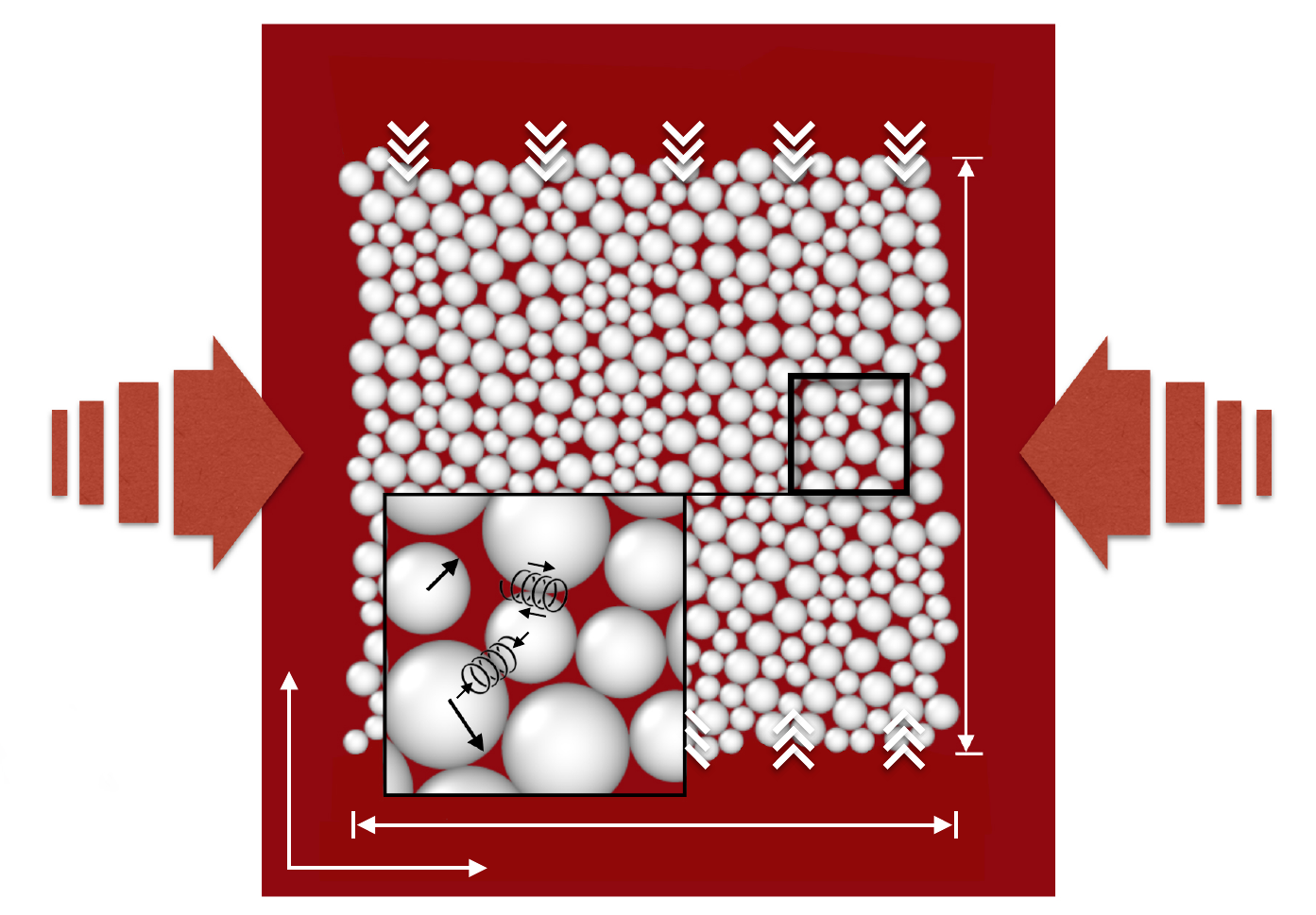}
			\put (15,35) {\color{white}$p_0$} 
			\put (82,35) {\color{white}$p_0$}
			\put (28,63) {\color{white}$\dot\epsilon_{yy}$} 
			\put (21,20) {\color{white}$y$} 
			\put (38,2.5) {\color{white}$x$} 
			\put (74.25,34) {\color{white}\begin{turn}{90}$L$\end{turn}}
			\put (49,3) {\color{white}\begin{turn}{0}$L$\end{turn}}
			\put (29,24) {\begin{turn}{45}\tiny$R_s$\end{turn}} 
			\put (31,15) {\begin{turn}{-40}\tiny$R_b$\end{turn}} 
			\put (39.5,18.5) {\begin{turn}{45}\tiny$k_n$\end{turn}} 
			\put (40,29) {\begin{turn}{-17}\tiny$k_t$\end{turn}} 
		\end{overpic}
	\caption{Bi-axial compression setup. The white discs (with radii $R_s$ and $R_b$) represent the bulk sample with size $L$. 
	The overlapping grains interact via linear springs $k_{n(t)}$ as sketched in the inset. The lateral arrows indicate the confining pressure $p_0$ regulated by the barostat. 
	The vertical arrows indicate the strain-controlled condition with a constant axial strain rate of $\dot\epsilon_{yy}$.}
	\label{fig:setUP}
	\end{center}
\end{figure}
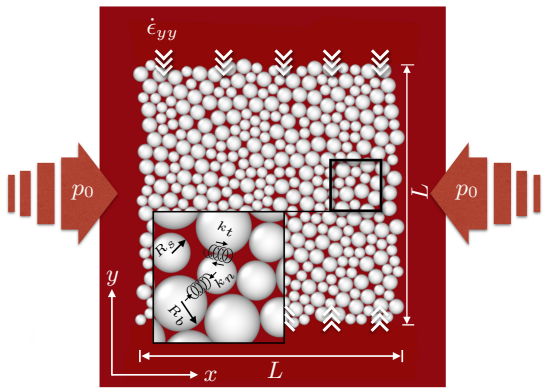
%
We performed a series of strain-controlled tests on samples with friction coefficient $\mu$ and initial pressure $p_0$.
The resulting load curves $\sigma=\frac{1}{2}(\sigma_{xx}-\sigma_{yy})$ with $\sigma_{xx}=-p_0$ against shear strain $\epsilon=\frac{1}{2}(\epsilon_{xx}-\epsilon_{yy})$ along with the evolution of volumetric strain $\epsilon_v=\epsilon_{xx}+\epsilon_{yy}$ are reported in Fig.~\ref{fig:stressStrain}(a) and (b).
A peak stress is typically developed in the frictional sample followed by a sheer reduction in strength as the loading continues.
We noted that the decay of stress tends to become more pronounced at higher friction and/or pressure (\emph{cf.} Supplementary Materials Sec.~\ref{sec:StressDiscontinuity}). 
In the absence of local friction, the stress rises monotonically toward its flowing state. 
The initial compaction phase --with $\epsilon_v<0$-- appears in both cases in view of the primary elastic response of a compressible medium to the imposed pressure.
The deformation is accompanied by substantial dilatancy in Fig.~\ref{fig:stressStrain}(b) prior to yielding, a common property of frictional granular media.
By contrast, Fig.~\ref{fig:stressStrain}(a) shows a well-established steady flow following the initial transient regime with almost no net volume change.  
\begin{figure}
	\begin{center}
		\begin{overpic}[width=8.6cm]{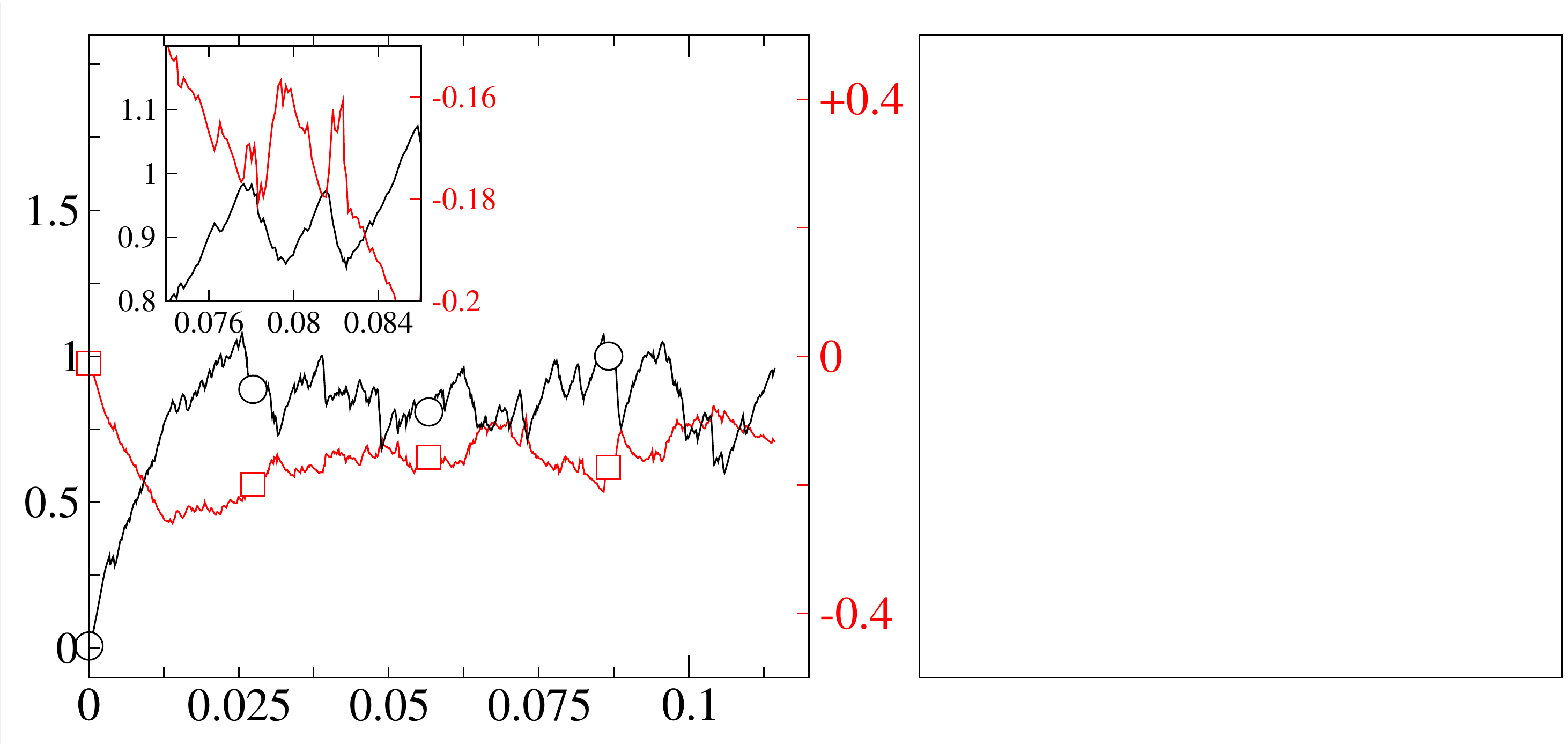}
			\put (27,7) {\small$\epsilon$} 
			\put (54,18) {\sffamily\setlength{\fboxsep}{0pt}\colorbox{white}{\strut\bfseries\textcolor{white}{\small\begin{turn}{90}{\color{red}{$\epsilon_{v}$\tiny${(\times 10^{-2})}$}}\end{turn}}}} 
			\put (-3,18) {\sffamily\setlength{\fboxsep}{0pt}\colorbox{white}{\strut\bfseries\textcolor{black}{\small\begin{turn}{90}{$\sigma$\tiny$(\times 10^{-3})$}\end{turn}}}} 
 		    \put(58.8,4.4){\includegraphics[height=3.5cm,width=3.5cm]{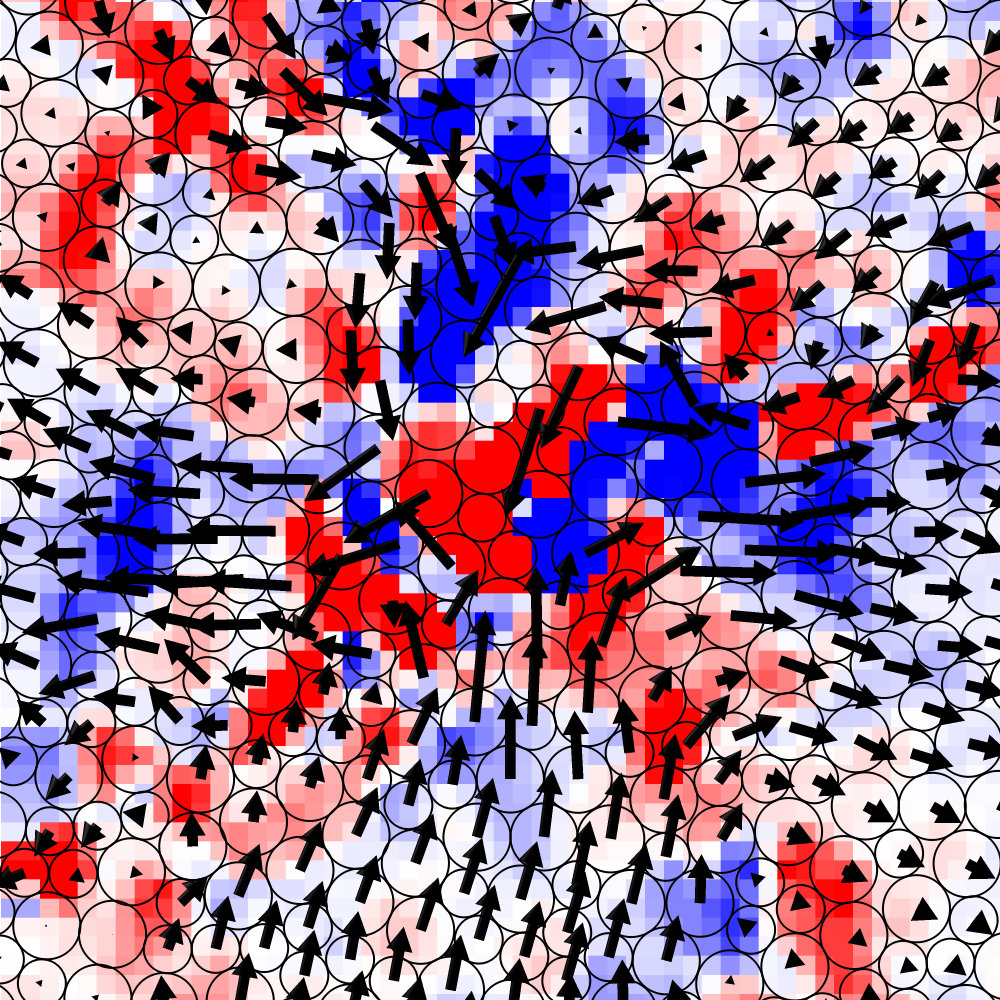}} 
			\put (60,7) {\sffamily\setlength{\fboxsep}{0pt}\colorbox{black}{\strut\bfseries\textcolor{white}{\small$(c)$}}} 
 		    \put(69.5,1){\includegraphics[height=0.15cm,width=1.7cm]{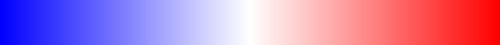}} 
			\put (65,1.5) {\color{black}\tiny$-2$} 
			\put (90,1.5) {\color{black}\tiny$2(\times10^{-2})$} 
			\put (7,7) {\sffamily\setlength{\fboxsep}{0pt}\colorbox{black}{\strut\bfseries\textcolor{white}{\small$(a)$}}} 
			\put (35,41) {\sffamily\setlength{\fboxsep}{0pt}\colorbox{black}{\strut\bfseries\textcolor{white}{\small~~$\mu=0$~~}}} 
			\begin{tikzpicture} 
				\coordinate (a) at (0,0); 
				\node[] at (a) {\tiny.};
				\coordinate (center2) at (2.75,1.7); 
				\coordinate (b) at ($ (center2) + 0.4*(1,0) $);
				\coordinate (c) at ($ (b) + 0.4*(0,1) $);
				\coordinate (d) at ($ (c) - 0.4*(1,0) $);
				\draw[line width=0.2mm,dashdotted] (center2) -- (b); 
				\draw[line width=0.2mm,dashdotted] (b) -- (c); 
				\draw[line width=0.2mm,dashdotted] (d) -- (c); 
				\draw[line width=0.2mm,dashdotted] (d) -- (center2); 
				\coordinate (e) at ($ (d) + .75*(-.85,1.5) $);
				\draw[line width=0.2mm,dashdotted] (d) -- (e); 
				\coordinate (a) at (0,0); 
				\node[] at (a) {\tiny.};				
				\coordinate (center3) at (2.75,1.15); 
				\coordinate (b) at ($ (center3) + 0.4*(1,0) $);
				\coordinate (c) at ($ (b) + 0.4*(0,1) $);
				\coordinate (d) at ($ (c) - 0.4*(1,0) $);
				\draw[line width=0.2mm,red,dashdotted] (center3) -- (b); 
				\draw[line width=0.2mm,red,dashdotted] (b) -- (c); 
				\draw[line width=0.2mm,red,dashdotted] (d) -- (c); 
				\draw[line width=0.2mm,red,dashdotted] (d) -- (center3); 
				\coordinate (e1) at ($ (d) + .77*(-1,1) $);
				\draw[line width=0.2mm,red,dashdotted] (d) -- (e1); 
			\end{tikzpicture}
					\end{overpic}
		\begin{overpic}[width=8.6cm]{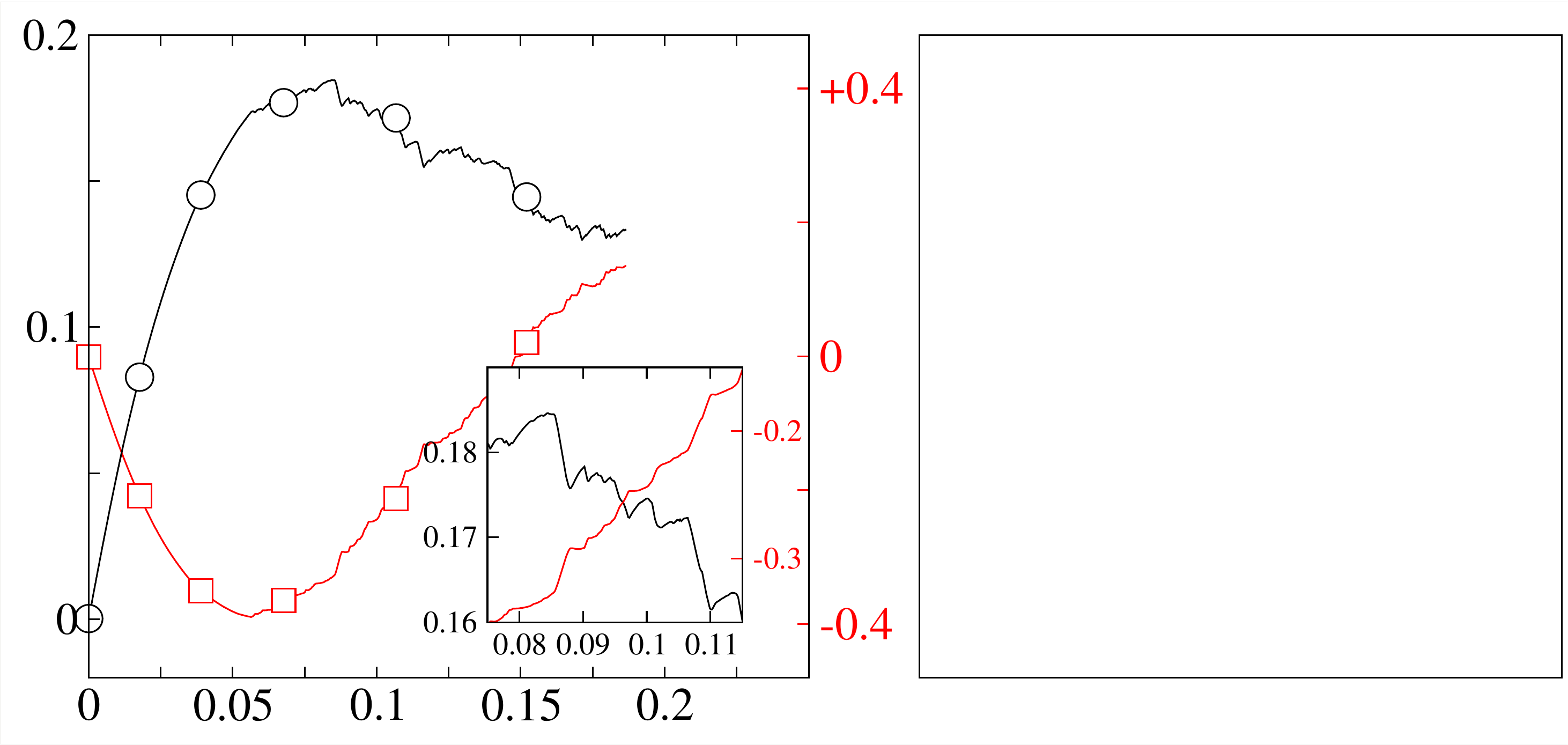}
			\put (54,18) {\sffamily\setlength{\fboxsep}{0pt}\colorbox{white}{\strut\bfseries\textcolor{white}{\small\begin{turn}{90}{\color{red}{$\epsilon_{v}$\tiny${(\times 10^{-1})}$}}\end{turn}}}} 
			\put (-3,18) {\sffamily\setlength{\fboxsep}{0pt}\colorbox{white}{\strut\bfseries\textcolor{black}{\small\begin{turn}{90}{$\sigma$\tiny$(\times 10^{-1})$}\end{turn}}}} 
			\put (35,41) {\sffamily\setlength{\fboxsep}{0pt}\colorbox{black}{\strut\bfseries\textcolor{white}{\small~$\mu=0.4$~}}} 
			\put (27,-3) {\small$\epsilon$} 
			\put (7,7) {\sffamily\setlength{\fboxsep}{0pt}\colorbox{black}{\strut\bfseries\textcolor{white}{\small$(b)$}}} 
 		    \put(58.8,4.4){\includegraphics[height=3.5cm,width=3.5cm]{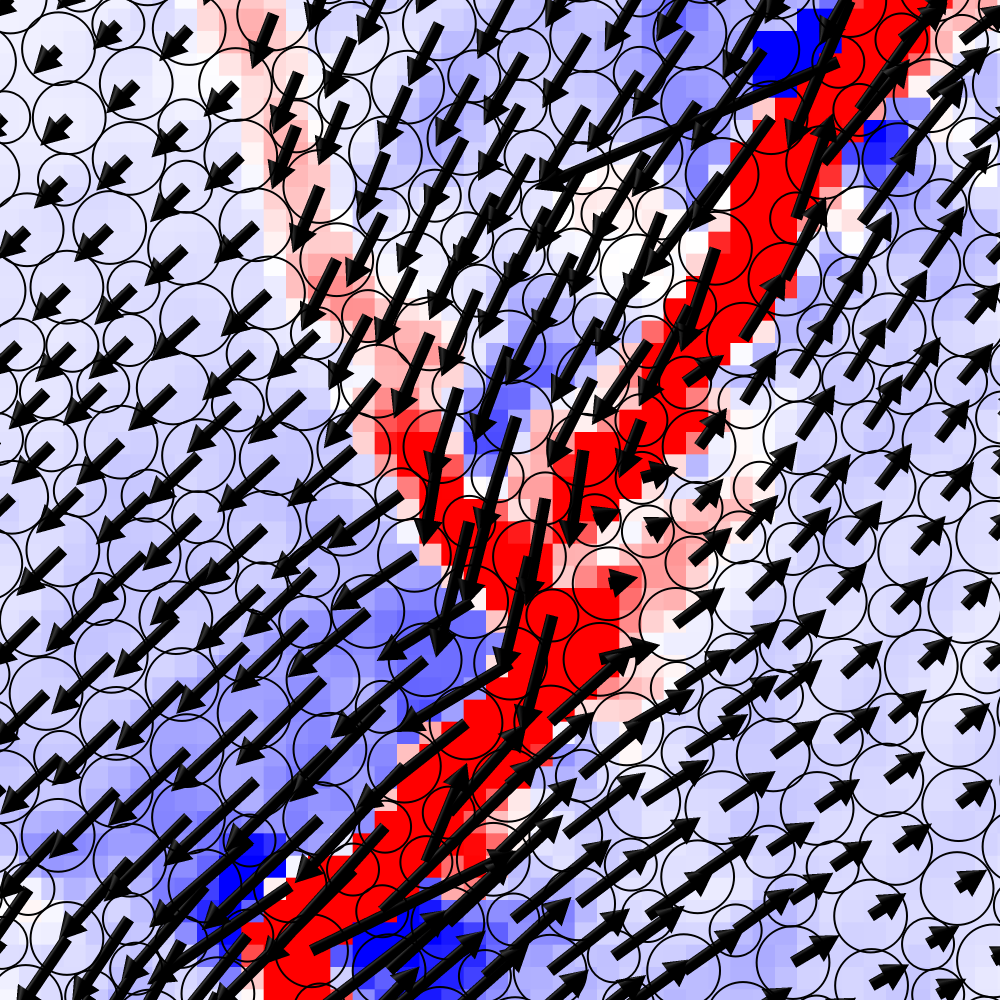}} 
			\put (60,7) {\sffamily\setlength{\fboxsep}{0pt}\colorbox{black}{\strut\bfseries\textcolor{white}{\small$(d)$}}} 
			\put(69.5,1){\includegraphics[height=0.15cm,width=1.7cm]{./colorBar.png}} 
			\put (65,1.5) {\color{black}\tiny$-2$} 
			\put (90,1.5) {\color{black}\tiny$2(\times10^{-2})$} 
			\begin{tikzpicture} 
				\coordinate (a) at (0,0); 
				\node[] at (a) {\tiny.};
				\coordinate (center2) at (1.5,3.25); 
				\coordinate (b) at ($ (center2) + 0.4*(1,0) $);
				\coordinate (c) at ($ (b) + 0.4*(0,1) $);
				\coordinate (d) at ($ (c) - 0.4*(1,0) $);
				\draw[line width=0.2mm,dashdotted] (center2) -- (b); 
				\draw[line width=0.2mm,dashdotted] (b) -- (c); 
				\draw[line width=0.2mm,dashdotted] (d) -- (c); 
				\draw[line width=0.2mm,dashdotted] (d) -- (center2); 
				\coordinate (e) at ($ (c) + .75*(.85,-2.25) $);
				\draw[line width=0.2mm,dashdotted] (b) -- (e); 
				\coordinate (a) at (0,0); 
				\node[] at (a) {\tiny.};				
				\coordinate (center3) at (1.5,.7); 
				\coordinate (b) at ($ (center3) + 0.4*(1,0) $);
				\coordinate (c) at ($ (b) + 0.4*(0,1) $);
				\coordinate (d) at ($ (c) - 0.4*(1,0) $);
				\draw[line width=0.2mm,red,dashdotted] (center3) -- (b); 
				\draw[line width=0.2mm,red,dashdotted] (b) -- (c); 
				\draw[line width=0.2mm,red,dashdotted] (d) -- (c); 
				\draw[line width=0.2mm,red,dashdotted] (d) -- (center3); 
				\coordinate (e1) at ($ (b) + .62*(1,0) $);
				\draw[line width=0.2mm,red,dashdotted] (b) -- (e1); 
			\end{tikzpicture}

		\end{overpic}
	\end{center}
	\caption{Results of strain-controlled bi-axial tests at (\textbf{a}) $\mu=0$ and (\textbf{b}) $\mu=0.4$ at $p_0=0.01$ and $N=80\times 80$. The bulk shear stress $\sigma$ and volumetric strain $\epsilon_v$ are denoted by different symbols and colors in each graph. The insets are close-up views of the main plots. Spatial maps of local shear strain $\epsilon(\vec{r})$ are also illustrated at (\textbf{c}) $\epsilon=0.08$ for the frictionless sample and (\textbf{d}) $\epsilon=0.1$ for the frictional specimen. Arrows mark displacements that occurred within the material over an unloading (stress drop) period.}
	\label{fig:stressStrain}
\end{figure}
Strong fluctuations are present in both samples at all times subsequent to the initial yielding. 
The stress dynamics is characterized by abrupt falloffs which are preceded by longer periods of stress build-up, an expected feature of amorphous structures as in the insets of Fig.~\ref{fig:stressStrain}(a) and (b).
Local shear strains $\epsilon(\vec{r})$, incurred during a typical stress drop, demonstrate the highly extended and anisotropic nature of bursts in the presence of frictional interactions as in Fig.~\ref{fig:stressStrain}(d).
The structure of the microscopic flow illustrated in Fig.~\ref{fig:stressStrain}(c) has the appearance of a classical Eshelby process.
Due to the quenched disorder, however, the actual response looks strongly scattered and it is only upon statistical averaging that fluctuations smooth out and the response will closely obey the proposed theoretical description.

\emph{Correlation Patterns: Point-like Eshelbys vs. Crack Arrays}-
We focus on the structure of fluctuations and quantify the associated spatial variations in a statistical manner. 
We probe the two-point correlation function between two different positions $\vec{r}{\hspace{1pt}^\prime}$ and $\vec{r}{\hspace{1pt}^{\prime\prime}}$ in space $C_\epsilon(\vec{r})\doteq\langle\epsilon(\vec{r}^{\hspace{1pt}\prime})\epsilon(\vec{r}{\hspace{1pt}^{\prime\prime}})\rangle$ with $\vec{r}=\vec{r}^{\hspace{1pt}\prime}-\vec{r}{\hspace{1pt}^{\prime\prime}}$.
The brackets $\langle...\rangle$ denote spatial averaging over different realizations.
$\epsilon(\vec{r}^{\hspace{1pt}\prime})$ corresponds to the shear deformation the material accommodates over the period of each stress drop like that of Fig.~\ref{fig:stressStrain}(c) or (d) \footnote{Statistically speaking, the kinematics of slip events incurred in the post-failure regime and under strain control nearly resembles those that occur in the stress-controlled protocol leading up to the failure point. We chose to probe the former group of events which take place at a higher frequency and, therefore, will give better statistics in terms of correlation features.}
Naturally, anisotropic strain patterns should induce distinct angular symmetries in the correlations as illustrated in Fig.~\ref{fig:crltns}(e) and (f). 
A typical four-fold structure is observed at $\mu=0$ with maximum lobes oriented along $45^\circ$ and $135^\circ$ which correspond to the maximum shear stress planes. 
The anisotropic part $C_{\epsilon}(\theta)=\langle C_\epsilon(\vec{r})\rangle_r$, the averaged correlation function over distance $r=|\vec{r}|$ along $\theta$, contains the quadropolar symmetry, as illustrated in Fig.~\ref{fig:crltns}(b).
The radial part $C_{\epsilon}(r)$ is displayed in Fig.~\ref{fig:crltns}(a) along four different orientations.
The power-law decay, over almost two decades of distance $r$, marks the non-local nature of fluctuations during the plastic flow.
The correlations fall off as $r^{-\nu}$ with the scaling exponent that seems to have modest angular dependence as reported in dense sheared glasses \cite{maloney2009anisotropic}.
We find $\nu\simeq1$ at $\theta=0^\circ({\color{green}\diamond}),90^\circ({\color{blue}\triangle})$ and $\nu\simeq2$ for $\theta=45^\circ({\color{black}\circ}), 135^\circ({\color{red}\triangle})$.   
The latter scaling may be understood on the basis of localized Eshelby events with long-range disturbance of the form $r^{-d}$ in the surrounding $d$-dimensional medium \cite{eshelby1957determination}.

\begin{figure}[t]
	\begin{center} 
		\begin{overpic}[width=8.6cm]{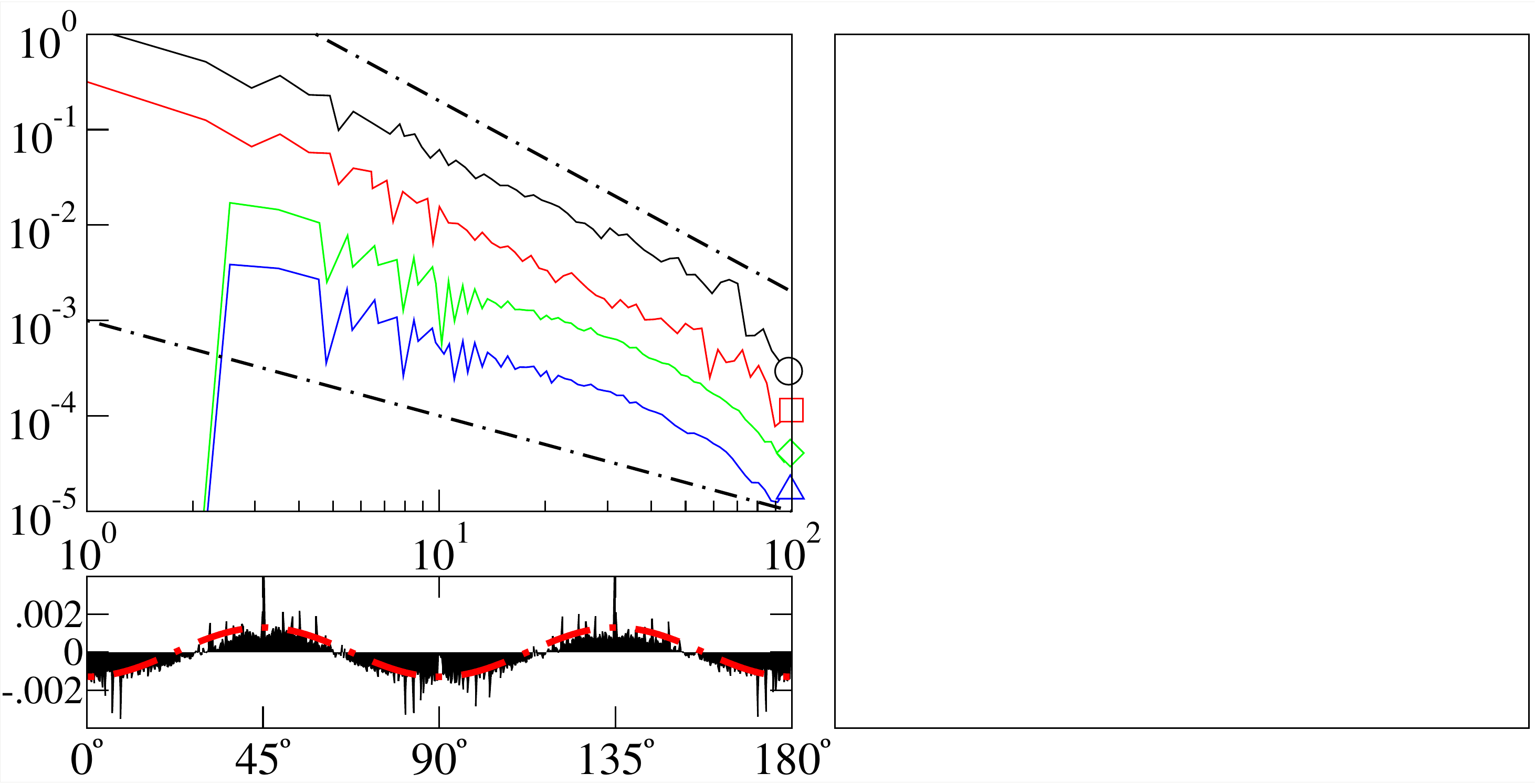}
			\put (28,20) {\sffamily\setlength{\fboxsep}{1pt}\colorbox{white}{\strut\bfseries\textcolor{black}{\small{{$r$}}}}} 
			\put (28,5) {\sffamily\setlength{\fboxsep}{0pt}\colorbox{white}{\strut\bfseries\textcolor{black}{\small{{$\theta$}}}}} 
			\put (-4,28) {\small\begin{turn}{90}{\small$C_{\epsilon}(r)$}\end{turn}} 
			\put (-4,4) {\small\begin{turn}{90}{\small$C_{\epsilon}(\theta)$}\end{turn}} 
			\put (8.65,45.1) {\tiny$45^\circ$} 
			\put (7.25,41.1) {\tiny\color{red}$135^\circ$} 
			\put (10,37.1) {\tiny\color{green}$0^\circ$} 
			\put (8.8,33.1) {\tiny\color{blue}$90^\circ$} 
 		    \put(54.9,4.1){\includegraphics[height=3.8cm,width=3.8cm]{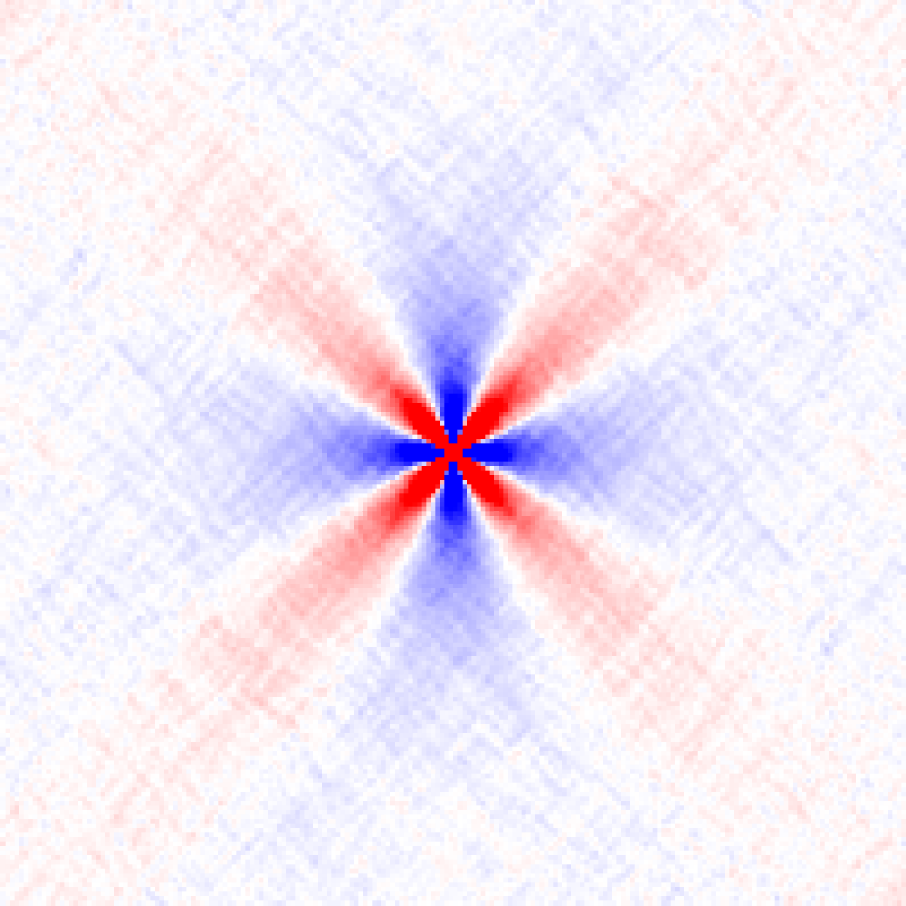}}
 		    \put(67.5,1){\includegraphics[height=0.15cm,width=1.7cm]{./colorBar.png}}
			\put (63,1.0) {\color{black}\tiny$-1$} 
			\put (88,1.0) {\color{black}\tiny$1(\times10^{-2})$} 
			\put (45.6,45) {\sffamily\setlength{\fboxsep}{0pt}\colorbox{black}{\strut\bfseries\textcolor{white}{\small$(a)$}}} 
			\put (45.6,10) {\sffamily\setlength{\fboxsep}{0pt}\colorbox{black}{\strut\bfseries\textcolor{white}{\small$(b)$}}} 
			\put (94,45) {\sffamily\setlength{\fboxsep}{0pt}\colorbox{black}{\strut\bfseries\textcolor{white}{\small$(e)$}}} 
			\begin{tikzpicture} 
				\coordinate (a) at (0,0); 
				\node[] at (a) {\tiny.};
				\coordinate (center) at (6.49,2.15); 
				\draw[dashdotted,thick] (center) -- ( $ (center) + 1.25*(1,1) $); 
				\draw[dashdotted,thick] (center) -- ( $ (center) + (1,0) $); 
				\draw ($(center)+(.2,0)$) arc(0:45:2mm); 
				\draw[fill=black] ($(center)+(.4,.05)$) rectangle ($(center)+(.4,.05)+(0.47,0.3)$); 
				\node[text=white] at ($(center)+(.64,.21)$) {\small{$45^\circ$}}; 
				\coordinate (center1) at (1.25,1.95); 
				\coordinate (b) at ($ (center1) + 3.9*0.2*(1,0) $);
				\coordinate (c) at ($ (center1) + 0.9*.2*(0,1) $); 
				\draw[line width=0.1mm] (center1) -- (b); 
				\draw[line width=0.1mm] (center1) -- (c); 
				\draw[line width=0.1mm] (b) -- (c); 
				\node at ($(center1)+(0.38,-0.1)$) {\tiny{1}}; 
				\node at ($(center1)+(-0.08,0.08)$) {\tiny{1}}; 
				\coordinate (center2) at (3.3,3.55); 
				\coordinate (b) at ($ (center2) - 5.3*.1*(1,0) $);
				\coordinate (c) at ($ (center2) - 2.8*.1*(0,1) $);
				\draw[line width=0.1mm] (center2) -- (b); 
				\draw[line width=0.1mm] (center2) -- (c); 
				\draw[line width=0.1mm] (b) -- (c); 
				\node at ($(center2)+(-0.25,0.1)$) {\tiny{1}}; 
				\node at ($(center2)+(.1,-0.1)$) {\tiny{2}}; 
				\coordinate (center2) at (7.6,.5); 
				\draw[thick] (center2) -- ( $ (center2) + 0.5*(1,0) $); 
				\draw[thick] ($ (center2) + 0.5*(1,0)-0.1*(0,1) $) -- ( $ (center2) + 0.5*(1,0)+0.1*(0,1) $); 
				\draw[thick] ($ (center2) -0.1*(0,1) $) -- ( $ (center2) +0.1*(0,1) $); 
				\node[] at ($(center2)+0.5*(1,0)+(-0.25,0.2)$) {\tiny{$50R_s$}}; 
			\end{tikzpicture}
		\end{overpic} 
		\begin{overpic}[width=8.6cm]{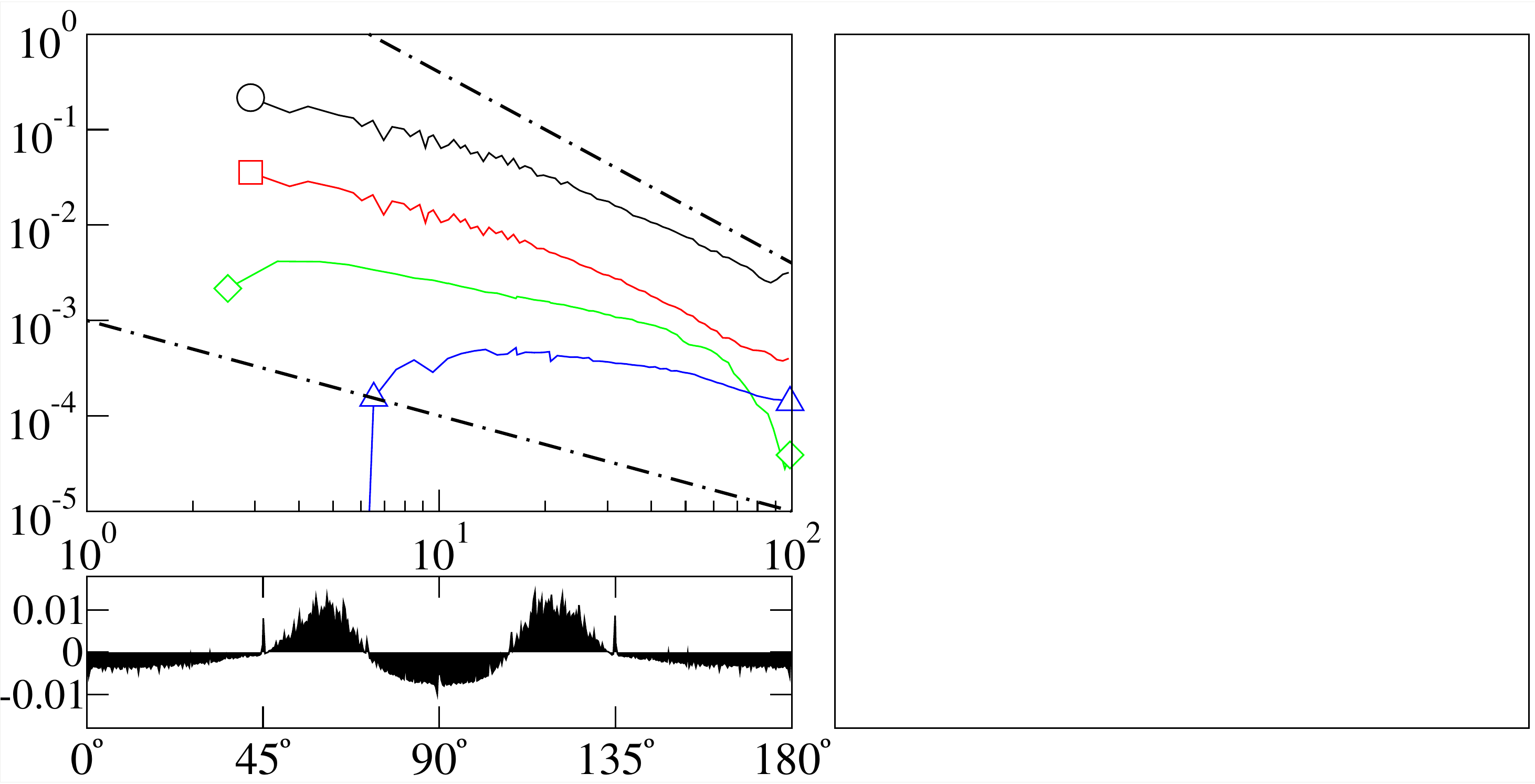}
			\put (28,20) {\sffamily\setlength{\fboxsep}{1pt}\colorbox{white}{\strut\bfseries\textcolor{black}{\small{{$r$}}}}} 
			\put (28,5) {\sffamily\setlength{\fboxsep}{0pt}\colorbox{white}{\strut\bfseries\textcolor{black}{\small{{$\theta$}}}}} 
			\put (-4,28) {\small\begin{turn}{90}{\small$C_{\epsilon}(r)$}\end{turn}} 
			\put (-4,4) {\small\begin{turn}{90}{\small$C_{\epsilon}(\theta)$}\end{turn}} 
			\put (10.65,44.5) {\tiny$60^\circ$} 
			\put (9.25,39.5) {\tiny\color{red}$120^\circ$} 
			\put (12,33.5) {\tiny\color{green}$0^\circ$} 
			\put (21,28.5) {\tiny\color{blue}$90^\circ$} 
 		    \put(54.9,4.1){\includegraphics[height=3.8cm,width=3.8cm]{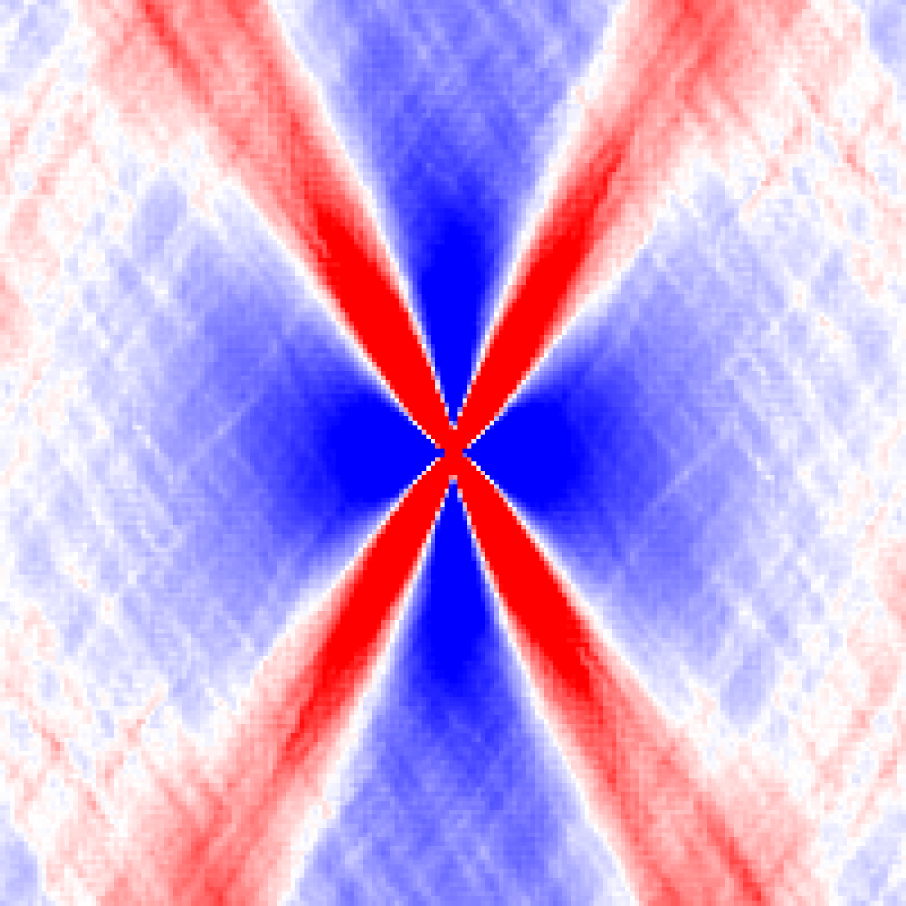}}
 		    \put(67.5,1){\includegraphics[height=0.15cm,width=1.7cm]{./colorBar.png}}
			\put (63,1.0) {\color{black}\tiny$-1$} 
			\put (88,1.0) {\color{black}\tiny$1(\times10^{-2})$} 
			\put (45.6,45) {\sffamily\setlength{\fboxsep}{0pt}\colorbox{black}{\strut\bfseries\textcolor{white}{\small$(c)$}}} 
			\put (45.6,10) {\sffamily\setlength{\fboxsep}{0pt}\colorbox{black}{\strut\bfseries\textcolor{white}{\small$(d)$}}} 
			\put (94,45) {\sffamily\setlength{\fboxsep}{0pt}\colorbox{black}{\strut\bfseries\textcolor{white}{\small$(f)$}}} 
			\begin{tikzpicture} 
				\coordinate (a) at (0,0); 
				\node[] at (a) {\tiny.}; 
				\coordinate (center) at (6.49,2.15); 
				\draw[dashdotted,thick] (center) -- ( $ (center) + 1.0*(1,1.765) $); 
				\draw[dashdotted,thick] (center) -- ( $ (center) + (1,0) $); 
				\draw ($(center)+(.2,0)$) arc(0:55:2mm); 
				\draw[fill=black] ($(center)+(.4,.05)$) rectangle ($(center)+(.4,.05)+(0.47,0.3)$); 
				\node[text=white] at ($(center)+(.64,.21)$) {\small{$60^\circ$}}; 
				\coordinate (center1) at (1.25,1.95); 
				\coordinate (b) at ($ (center1) + 3.9*0.2*(1,0) $);
				\coordinate (c) at ($ (center1) + 0.9*.2*(0,1) $); 
				\draw[line width=0.1mm] (center1) -- (b); 
				\draw[line width=0.1mm] (center1) -- (c); 
				\draw[line width=0.1mm] (b) -- (c); 
				\node at ($(center1)+(0.38,-0.1)$) {\tiny{1}}; 
				\node at ($(center1)+(-0.08,0.08)$) {\tiny{1}}; 
				\coordinate (center2) at (3.3,3.72); 
				\coordinate (b) at ($ (center2) - 5.3*.1*(1,0) $);
				\coordinate (c) at ($ (center2) - 2.8*.1*(0,1) $);
				\draw[line width=0.1mm] (center2) -- (b); 
				\draw[line width=0.1mm] (center2) -- (c); 
				\draw[line width=0.1mm] (b) -- (c); 
				\node at ($(center2)+(-0.25,0.1)$) {\tiny{1}}; 
				\node at ($(center2)+(.1,-0.1)$) {\tiny{2}}; 
				\coordinate (center2) at (7.6,.5); 
				\draw[thick] (center2) -- ( $ (center2) + 0.5*(1,0) $); 
				\draw[thick] ($ (center2) + 0.5*(1,0)-0.1*(0,1) $) -- ( $ (center2) + 0.5*(1,0)+0.1*(0,1) $); 
				\draw[thick] ($ (center2) -0.1*(0,1) $) -- ( $ (center2) +0.1*(0,1) $); 
				\node[] at ($(center2)+0.5*(1,0)+(-0.25,0.2)$) {\tiny{$50R_s$}}; 
			\end{tikzpicture}

		\end{overpic}

	\end{center}
	\caption{Strain correlations $C_{\epsilon}(\vec{r})$ at $p_0=0.01$ for (\textbf{top}) $\mu=0$  and (\textbf{bottom}) $\mu=0.4$. (\textbf{a},\textbf{c}) The radial part of the correlation function $C_{\epsilon}(r)$ along four different orientations as indicated in the graphs. The curves are shifted vertically for the sake of clarity. The dashed lines are guides to the power laws with the corresponding exponents. Note that $-C_{\epsilon}(r)$ is plotted at $\theta=0^{\circ}$ and $90^{\circ}$.  (\textbf{b},\textbf{d}) The anisotropic part $C_{\epsilon}(\theta)$ vs. $\theta$. The dashed (red) curve marks $C_{\epsilon}(\theta)\propto-\text{cos}~4\theta$. (\textbf{e},\textbf{f}) Correlation maps. The scale bars denote the scale of each map. The dashed lines indicate maximal correlation directions.} 
	\label{fig:crltns}
\end{figure}
In the presence of friction, the quadro-polar shape is distorted as in Fig.~\ref{fig:crltns}(d) with the correlation profile growing sharply toward its maxima.
The maximal correlation angles differ from the canonical $45^\circ$ and $135^\circ$ becoming tilted toward planes with lower normal stresses, an expected behavior of pressure-sensitive materials.  
The characteristic angle $\theta\simeq60^\circ$, indicated by the dashed line in Fig.~\ref{fig:crltns}(f), seems to be in disagreement with the Mohr-Coulomb orientation $\theta_\text{MC}=45^\circ+\frac{\phi}{2}$ with $\phi=\text{tan}^{-1}\mu\simeq22^\circ$.
The latter orientation is traditionally treated as a macroscopic failure angle in granular aggregates and brittle rocks \cite{craig2013soil} and, as we evidenced in Supplementary Materials Sec.~\ref{sec:MohrCoulomb}, $\theta_\text{MC}$ somewhat underestimates the preferential inclination of strain fluctuations.
Figure~\ref{fig:crltns}(c) examines the spatial decay of $C_{\epsilon}(r)$ along four orientations.
At $\theta=0^\circ({\color{green}\diamond}), 90^\circ({\color{blue}\triangle})$, within the negatively correlated sectors, fluctuations become nearly flat having weak variations with $r$.   
At $\theta=60^\circ(\circ), \theta=120^\circ({\color{red}\Box})$, the data falls off as $r^{-1}$ at small and intermediate scales and more steeply as $r^{-2}$ at larger distances.
We conjecture that the former scaling signifies the role of \emph{near-field} contributions in the presence of finite-sized slip zones which can be  treated as Eshelby features in the \emph{far-field}, giving grounds for the latter power-law decay. 
The non-remote approximation is, however, at odds with the classical $\sigma_{\infty}{\sqrt \frac{a}{r}}$ stress singularity \emph{near} the crack tip of size $a$ loaded remotely by uniform shear $\sigma_{\infty}$ \cite{scholz2002mechanics}. 
The discrepancy is despite the close resemblance, in a topological sense, between the incident shear \emph{faults} and Mode II fractures.
In fracture mechanics, the stress intensity is mediated by the geometry of shear cracks and elasticity of the embedding solid whereas 
the near-field divergence in our data is in parts due to the progressive coalescence of interacting Eshelby events leading to the formation of \emph{fractal} units with dimension $d_f\simeq 1$.
Therefore, $\nu=d-d_f$ taking into account the morphology of the embedded elements which are quite elongated in space and, as a result,  induce longer range perturbations of the form $r^{-(d-d_f)}$ within the solid matrix.

The geometric structure of the deformation patterns reflects the mesoscopic character of the frictional flow.
That is, yielding initiates locally via scattered small-scale cracks that propagate as the stored stress (or energy) at the crack tip reaches some local thresholds, \emph{i.e.} critical intensity factor or toughness $K_c$ in the context of fracture mechanics.
Due to random fluctuations in fracture toughness and heterogeneities, the propagation halts after the crack front hits rigid regions in space.
The radiated energy will subsequently induce long-range stress fluctuations which decay as $\frac{1}{r}$ in our model and will activate vulnerable cracks in the surrounding medium. 
This avalanche-like collective dynamics preferentially occurs along characteristic angles which result from long-range elasticity coupled with local friction law \cite{karimi2018correlation}.

%
\emph{Damage accumulation vs. mechanical healing}-
Our key observation is that the addition of friction law at the grain scale alters the constitutive behavior from ductile flow to pressure-sensitive brittle deformation. 
Apart from fluctuation patterns, the transition is manifested in terms of the \emph{mean} response prior to the failure point.
In Fig.~\ref{fig:stressStrain}(b), the stress response is accompanied by remarkable \emph{softening} --an absent feature in Fig.~\ref{fig:stressStrain}(a)-- which is spatially associated with nonaffine inelastic transformations within the frictional medium.
This mechanism is comparable to the progressive damage with nucleation of micro-cracks that will effectively deteriorate elastic properties in brittle materials \cite{vu2018compressive,renard2017microscale}.
In the framework of continuum damage \cite{kachanov1958time, lemaitre1994mechanics}, this effect is quantified via the damage function $D$ relating the (shear) modulus of the undamaged state to that of a previously-loaded stress-free sample.

To validate the damage hypothesis, we performed a number of oscillatory shear tests which enables the determination of the bulk shear modulus $G$ at multiple pre-failure stress levels $\sigma<\sigma_f$ (Supplementary Materials Sec.~\ref{sec:DamageAnalysis}).
The damage index is defined as $D\doteq 1-\frac{G}{G_0}$ with $G_0=G|_{\sigma\rightarrow 0}$ denoting the sample rigidity in the pre-damage regime. 
Figure~\ref{fig:damage5} displays the evolution of $G$ and $D$ with $\sigma$ and $\Delta=1-\frac{\sigma}{\sigma_f}$, respectively. 
The plateau regime in Fig.~\ref{fig:damage5}(a) suggests that, in a purely plastic flow, the material heals itself with immediate recovery of the initial elastic property and almost no memory of its previously flowing state.
This contrasts with the progressive damage scenario corresponding to the frictional aggregate in Fig.~\ref{fig:damage5}(b) which signs the presence of long-term memory that accompanies the pre-failure deformation even though \emph{no} irreversible bond breaking occurs at the particle scale.

\emph{Failure Transition: sharp vs. smooth}-
The critical fluctuations illustrated in Fig.~\ref{fig:unloading} demonstrate the highly intermittent nature of pre-yielding dynamics.
The macroscopic observable, $S$, probed here physically represents the accelerating energy release that accompanies the deformation.
$\sigma_f$ denotes the failure stress in stress-controlled tests that were performed so as to track accurately the evolution of this quantity (see Supplementary Materials Sec.~\ref{sec:SimulationsAndProtocols} for the definition of $S$).
Our focus will be on the failure prediction context and the evolution of the relevant precursory index in our model, \emph{i.e.} the maximal avalanche size $S_\text{max}$ and the corresponding mean value $\langle S\rangle$ \cite{girard2010failure,girard2012damage}. %
\begin{figure}[t]
	\begin{center}
		\begin{overpic}[width=8.6cm]{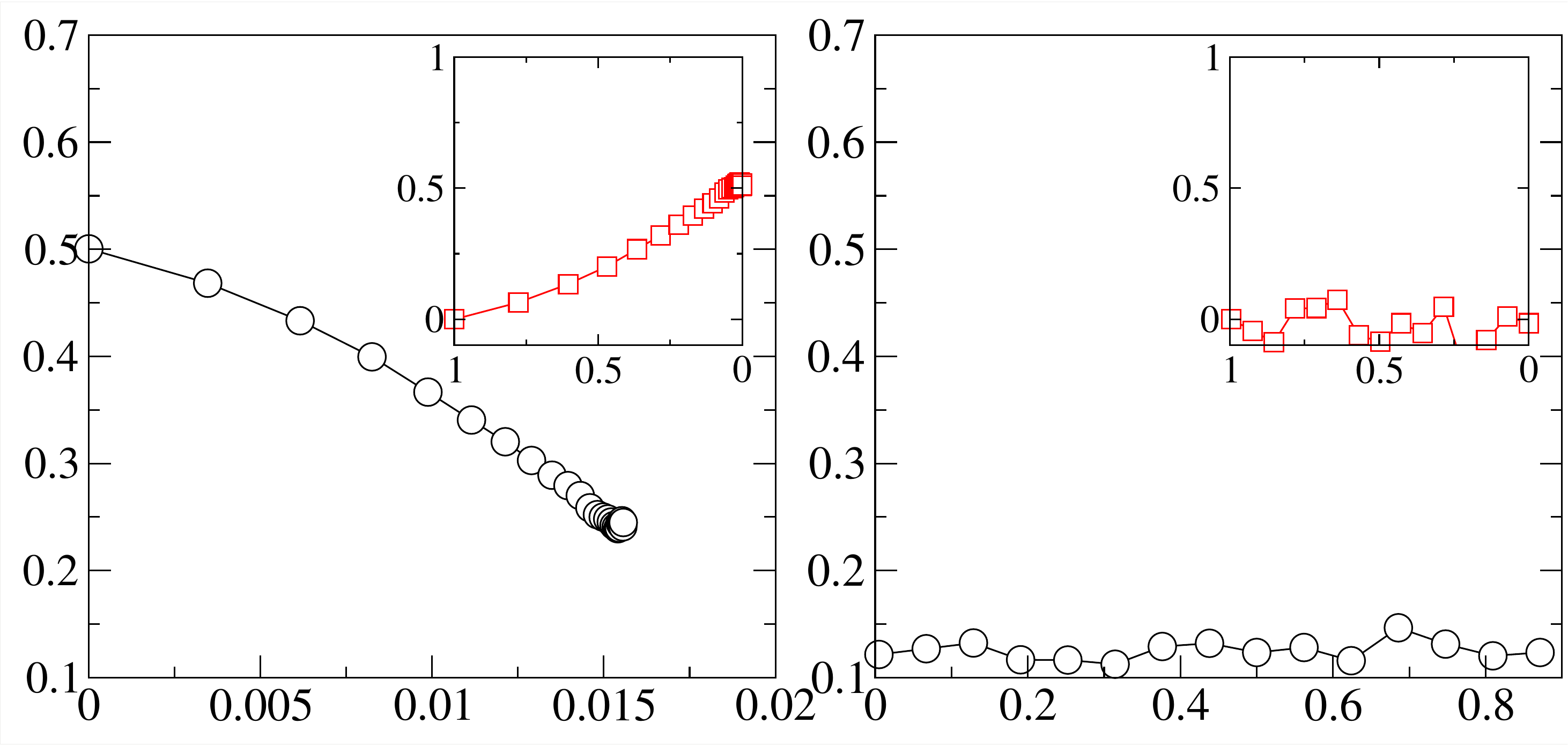}
            \put (26,-2) {$\sigma$}
            \put (76,-2) {$\sigma$\tiny$(\times10^{-3})$} 
            \put (87,20) {\tiny$\Delta$} 
            \put (37,20) {\tiny$\Delta$} 
            \put (-4,24) {\sffamily\setlength{\fboxsep}{0pt}\colorbox{white}{\strut\bfseries\textcolor{black}{\begin{turn}{90}{$G$}\end{turn}}}} 
            \put (101,24) {\sffamily\setlength{\fboxsep}{0pt}{\strut\bfseries\textcolor{black}{\begin{turn}{90}{$G$}\end{turn}}}} 
            \put (72,34) {\sffamily\setlength{\fboxsep}{0pt}{\strut\bfseries\textcolor{black}{\tiny\begin{turn}{90}{$D$}\end{turn}}}} 
            \put (22,34) {\sffamily\setlength{\fboxsep}{0pt}{\strut\bfseries\textcolor{black}{\tiny\begin{turn}{90}{$D$}\end{turn}}}} 
            \put (7,41.5) {\sffamily\setlength{\fboxsep}{0pt}\colorbox{black}{\strut\bfseries\textcolor{white}{\small$(a)$}}} 
            \put (57,41.5) {\sffamily\setlength{\fboxsep}{0pt}\colorbox{black}{\strut\bfseries\textcolor{white}{\small$(b)$}}} 
            		\end{overpic}
		\end{center}
	\caption{Evolution of the shear modulus $G$ with the applied stress $\sigma$ at $p=10^{-2}$ and $N=80\times 80$ for (\textbf{a}) $\mu=0$ and (\textbf{b}) $\mu=0.4$. The insets plot the damage index $D$ vs. $\Delta$.}
	\label{fig:damage5}
\end{figure}
 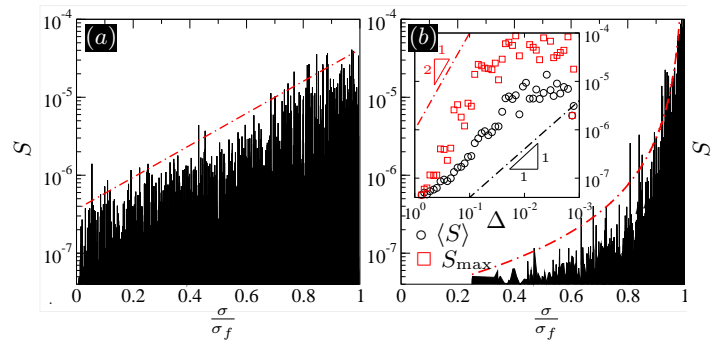
\begin{figure}[t]
	\begin{center} 
		\begin{overpic}[width=8.6cm]{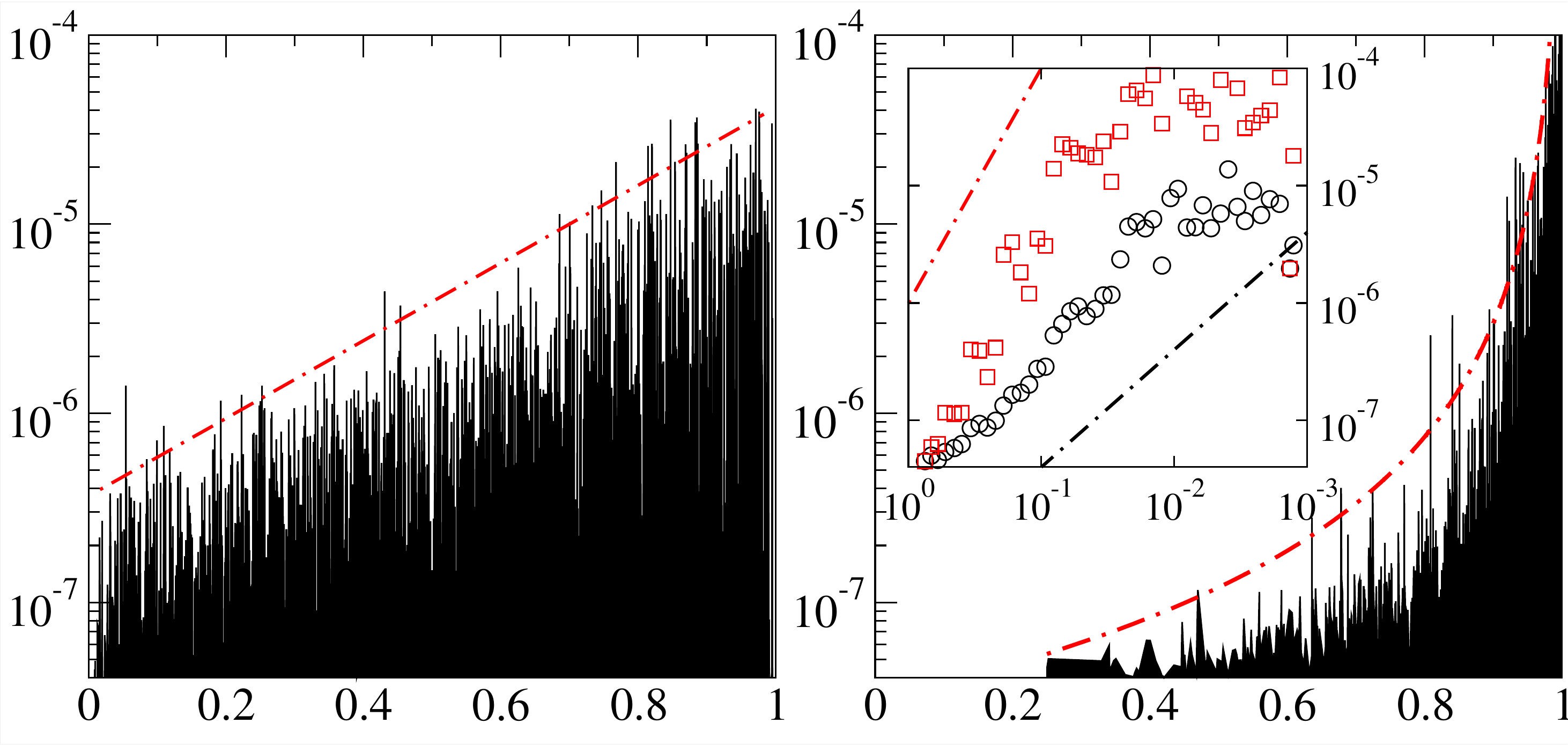}
             \put (26,-2) {$\frac{\sigma}{\sigma_f}$} 
             \put (69,13) {\small$\Delta$} 
             \put (58,11) {$\circ~\langle S\rangle$} 
             \put (58,7) {${\color{red}\Box}~S_\text{max}$} 
             \put (92,16) {$\mu$} 
             \put (-4,24) {\sffamily\setlength{\fboxsep}{0pt}\colorbox{white}{\strut\bfseries\textcolor{black}{\begin{turn}{90}{$S$}\end{turn}}}} 
             \put (101,24) {\sffamily\setlength{\fboxsep}{0pt}{\strut\bfseries\textcolor{black}{\begin{turn}{90}{$S$}\end{turn}}}} 
             \put (76,-2) {$\frac{\sigma}{\sigma_f}$} 
             \put (7,41.5) {\sffamily\setlength{\fboxsep}{0pt}\colorbox{black}{\strut\bfseries\textcolor{white}{\small$(a)$}}} 
             \put (57,41.5) {\sffamily\setlength{\fboxsep}{0pt}\colorbox{black}{\strut\bfseries\textcolor{white}{\small$(b)$}}} 
			\begin{tikzpicture}
                 \coordinate (a) at (0,0); 
                 \node[] at (a) {\tiny.};
                 \coordinate (center1) at (6.1,1.78); 
                 \coordinate (b) at ($ (center1) + .375*(1,0) $);
                 \coordinate (c) at ($ (b) + 0.345*(0,1) $); 
                 \draw[line width=0.1mm] (center1) -- (b); 
                 \draw[line width=0.1mm] (center1) -- (c); 
                 \draw[line width=0.1mm] (b) -- (c); 
                 \node at ($(center1)+(0.19,-0.1)$) {\tiny{1}}; 
                 \node at ($(center1)+(.48,0.15)$) {\tiny{1}}; 
                 \coordinate (center1) at (5.1,2.9); 
                 \coordinate (b) at ($ (center1) + .345*(0,1) $);
                 \coordinate (c) at ($ (b) + 0.2*(1,0) $); 
                 \draw[line width=0.1mm,red] (center1) -- (b); 
                 \draw[line width=0.1mm,red] (center1) -- (c); 
                 \draw[line width=0.1mm,red] (b) -- (c); 
                 \node at ($(center1)+(0.1,0.47)$) {\color{red}\tiny{1}}; 
                 \node at ($(center1)+(-.09,0.20)$) {\color{red}\tiny{2}}; 

             \end{tikzpicture}
             		\end{overpic}
		\end{center}
	\caption{The avalanche size $S$ in a stress-controlled setup plotted against $\frac{\sigma}{\sigma_f}$ prior to the ultimate failure at $p=10^{-2}$ and $N=80\times80$ with (\textbf{a}) $\mu=0$ and (\textbf{b}) $\mu=0.4$. The inset illustrates the evolution of $\langle S\rangle$ and $S_\text{max}$ with $\Delta=1-\frac{\sigma}{\sigma_f}$ on log-log scale. The dash-dotted curves in the inset are guides to the power law. The red curves in the main plots indicate $S_\text{max}\propto\text{exp}(-\frac{\Delta}{\Delta^*})$ with $\Delta^*$ being a constant in (a) and the inverse squared divergence $S_\text{max}\propto\frac{1}{\Delta^2}$ in (b).}
	\label{fig:unloading}
\end{figure}
The statistics were obtained independently out of over 100 samples at $p_0=0.01$.
In Fig.~\ref{fig:unloading}(a), the fluctuations seem to grow exponentially in size for $\mu=0$, but accelerate at $\mu=0.4$ with a form that closely matches an algebraic divergence as shown in Fig.~\ref{fig:unloading}(b). 
The inset displays the evolution of $\langle S\rangle$ and $S_\text{max}$ as a function of $\Delta$.
The fluctuations correspond with slip events that occur over a broad range of scales limited by the distance to the criticality and, naturally, the physical system size. 
The latter, in fact, corresponds to the plateau region as $\Delta\rightarrow 0$.
In the early stages of loading, $0.5\le\Delta<1$, no or very few precursory signals are detected at the higher friction value which makes the ensuing transition rather steep and, therefore, more brittle-like as the failure approaches.
This is at odds with the rather smooth evolution of the order parameter in frictionless samples which sets in right at the outset of the deformation and indicates plastic shearing.

\emph{Conclusions}- In this work, we have taken a purely microscopic viewpoint of shear deformation in granular aggregates by considering the dynamics of local slip events. 
At the scale of individual constituents, frictionless grains relax internal stress via localized Eshelby-type modes which is quite distinguishable from abrupt shear faulting in the presence of friction.
The disparity between these two elementary mechanism has a direct implication on the nature of macroscopic response; while the former mechanism leads to a bulk plastic flow, the latter feature may indicate a brittle failure transition.

Based upon this kinematic description, we have quantified the spatial strain correlation patterns that incurred during relaxation events corresponding to each process.
In one case, the angular relations and spatial decay of correlations conform to the expectations of an Eshelby-based process, \emph{i.e.} four-fold angular symmetry with ductile shearing that occurs on planes of maximum shear stress. 
In the context of frictional flow, the zone of strain localization did not look to be aligned with the Coulomb orientation. 
Besides, linear elastic fracture mechanics (LEFM) would lead us to expect a different scaling for the spatial variations of the near-field strain patterns for a Mode $\text{II}$ crack.
Fineberg \emph{et al.} recently showed that frictional interfaces can be considered as shear cracks as well, at least in terms of elastic strains with $\frac{1}{\sqrt{r}}$ singularity \cite{svetlizky2014classical}. 
This prediction does not seem to be admissible by our data where arrays of coordinated Eshelby features lead to $\frac{1}{r}$ scaling behavior.

The critical failure transition observed in the presence of friction may be attributed to progressive damage mechanism and associated weakening that is a common deformation process in pressure-sensitive materials and often results in catastrophic failure events \cite{vu2018compressive}.
In the absence of friction, on the other hand, the suppression of criticality can be interpreted in the context of mechanical \emph{healing}\cite{baro2018experimental,renard2017microscale,weiss2016cohesion}, a prevalent mechanism in ductile shear flows with microscopic constitutive elements that rearrange continually.
In fact, the failure dynamics in the latter case lacks a unique critical point as evidenced by the non-singular exponential-like acceleration of the radiated energy.
Similar observations were made in the context of crystal plasticity and dislocation avalanches \cite{ispanovity2014avalanches,lehtinen2016glassy} where the non-critical dynamics was attributed to the formation of dislocation junctions and associated strain hardening.

\emph{Acknowledgment-} We acknowledge financial support from the ANR grant ANR-16-CE30-0022-Relfi. We would also like to acknowledge useful discussions with A. Amon and J. Crassous about our results. We appreciate A. Roudgar Amoli's help with the graphical sketches.     
\newpage
\bibliography{ref}
\begin{center}
\large \bf Supplementary Materials
\end{center}
\maketitle
\setcounter{figure}{0}
\section{Simulations and protocols}\label{sec:SimulationsAndProtocols} We used bi-disperse packings of two dimensional grains with radii $R_s$ and $R_b$.
We set ${R_b}/{R_s}=1.4$ and ${N_b}/{N_s}=1$ where $N_{b(s)}$ denotes the number of grains in each species.
The $i\text{-th}$ and $j\text{-th}$ particles with position vectors $\vec{r}_i$, $\vec{r}_j$ and rotations $\theta_i$, $\theta_j$ may interact with each other when the overlap $\delta=R_i+R_j-|\vec{r}_i-\vec{r}_j|>0$. 
The normal and frictional contact forces are 
\begin{eqnarray}\label{eq:dynamics}
\vec{f}_n&=&-k_n~\delta~\vec{e}_n, \nonumber \\
\vec{f}_t&=&-k_t~\Delta u^{ij}_t~\vec{e}_t,
\end{eqnarray}
with the unit normal and tangential vectors $\vec{e}_n=(\vec{r}_i-\vec{r}_j)/|\vec{r}_i-\vec{r}_j|$ and $\vec{e}_t$.   
Here $k_n$ and $k_t$ are the normal and tangential spring constants, respectively.
The relative tangential displacement between the two grains in contact is computed as $\Delta u^{ij}_t=\int_{t^\prime}v_t~dt^\prime$ with the tangential components of the relative velocity $v_t=(\vec{\dot{r}}_i-\vec{\dot{r}}_j).\vec{e}_t+R_i\dot\theta_i+R_j\dot\theta_j$.

In the following, we will assume that the shear threshold can be expressed as a linear function of the normal contact force, in accordance with Coulomb's law.
Therefore, the distance to local failure at any contact point may be expressed 
by\begin{equation}
f_y=|\vec{f}_t|-\mu |\vec{f_n}|,
\end{equation} 
where $\mu$ is the microscopic friction coefficient. 
Below the failure limit, $f_y < 0$, the tangential force grows according to Eq.~\ref{eq:dynamics}.
Upon local yielding at $f_y=0$, the contact becomes fully plastic with the tangential force held constant until the grains lose contact.
A linear drag force $\vec{f}_\text{vis}=-m~\tau^{-1}_d~\vec{\dot{r}}$ is applied on each grain with mass $m$ and dissipation time-scale $\tau_d$. 
On top of that, a normal and tangential pair drag term $\vec{f}^\text{drag}_{n(t)}=-m~\gamma_{n(t)}~v_{n(t)}~\vec{e}_{n(t)}$ with $v_{n}=(\vec{\dot{r}}_i-\vec{\dot{r}}_j).\vec{e}_{n}$ was included with corresponding rates $\gamma_{n(t)}$.
The rate unit (inverse timescale) is set by the vibrational frequency $\omega^2_{n(t)}=k_{n(t)}/m$.
Newton's equations of motion were solved in LAMMPS \cite{plimpton1995fast} 
\begin{eqnarray}
m_i\vec{\ddot{u}}_i&=&\vec{f}_n+\vec{f}_t+\vec{f}_\text{vis}+\vec{f}^\text{drag}_{n}+\vec{f}^\text{drag}_{t}, \nonumber \\
m_iR_i^2\ddot\theta_i \vec{e}_z&=& R_i~\vec{e}_n\times(\vec{f}_t+\vec{f}^\text{drag}_t).
\end{eqnarray}
We also set the discretization time $\Delta t=0.05~\omega_n^{-1}$ and ${k_n}/{k_t} = 2$. 
An overdamped dynamics was imposed by setting a high value of the damping rates $\tau_d^{-1}$ and $\gamma_{n(t)}$ (in comparison with the vibrational frequency).
\begin{figure}[t]
	\begin{center} 
		\begin{overpic}[width=8.6cm,]{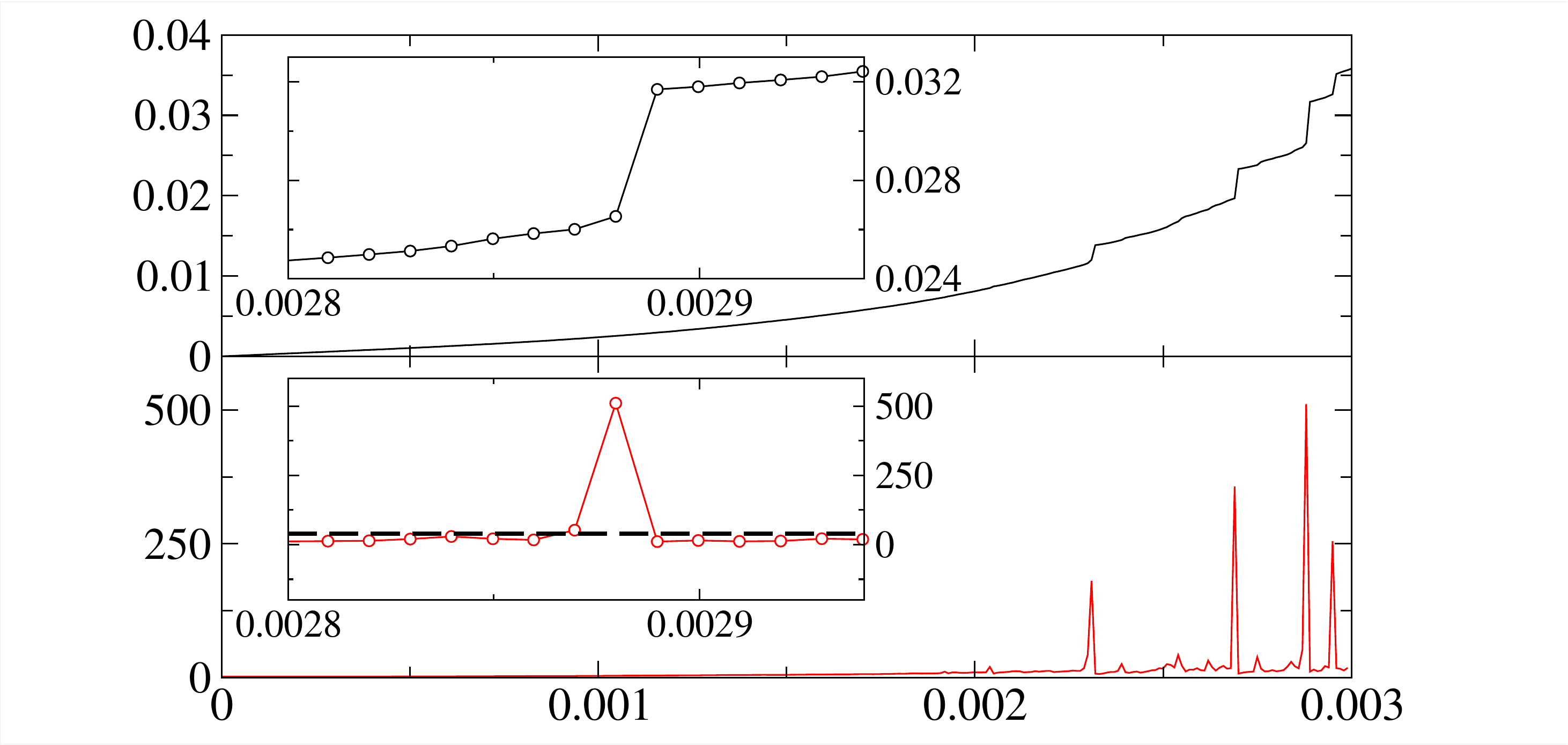}
             \put (50,-2) {$\sigma$} 
             \put (23,15) {$\dot\epsilon_c$} 
             \put (37,11) {\tiny\color{red}$\Delta\sigma$} 
             \put (36,6.7) {\tiny\color{red}$\sigma_i$} 
             \put (5,15) {\sffamily\setlength{\fboxsep}{0pt}\colorbox{white}{\strut\bfseries\textcolor{black}{\begin{turn}{90}{$\dot\epsilon$}\end{turn}}}} 
             \put (5,35) {\sffamily\setlength{\fboxsep}{0pt}\colorbox{white}{\strut\bfseries\textcolor{black}{\begin{turn}{90}{$\epsilon$}\end{turn}}}} 
             \put (33,37) {\sffamily\setlength{\fboxsep}{0pt}\colorbox{white}{\strut\bfseries\textcolor{red}{\begin{turn}{90}{$S$}\end{turn}}}} 
             \put (87,42.3) {\sffamily\setlength{\fboxsep}{0pt}\colorbox{black}{\strut\bfseries\textcolor{white}{\small$(a)$}}} 
             \put (87,21.9) {\sffamily\setlength{\fboxsep}{0pt}\colorbox{black}{\strut\bfseries\textcolor{white}{\small$(b)$}}} 
             \begin{tikzpicture}
                 \coordinate (a) at (0,0); 
                 \node[] at (a) {\tiny.};
			     \coordinate (center2) at (3.0,.63); 
                 \coordinate (b) at ($ (center2) + .05*(0,1) $); 
                 \coordinate (c) at ($ (center2) - .05*(0,1) $);
                 \draw[red][line width=0.1mm] (c) -- (b); 
			     \coordinate (center2) at (3.41,.85); 
                 \coordinate (b) at ($ (center2) + .1*(0,1) $); 
                 \coordinate (c) at ($ (center2) - .1*(0,1) $);
                 \draw[red][line width=0.1mm] (c) -- (b); 
                 \coordinate (d) at ($ (center2) + .2*(1,0) $); 
                 \draw[->,>=stealth,red][line width=0.1mm] (d) -- (center2); 
			     \coordinate (center2) at (3.0,.85); 
                 \coordinate (b) at ($ (center2) + .1*(0,1) $); 
                 \coordinate (c) at ($ (center2) - .1*(0,1) $);
                 \draw[red][line width=0.1mm] (c) -- (b); 
                 \coordinate (d) at ($ (center2) - .2*(1,0) $); 
                 \draw[->,>=stealth,red][line width=0.1mm] (d) -- (center2); 
			     \coordinate (center2) at (2.99,3.5); 
                 \coordinate (b) at ($ (center2) + .1*(1,0) $); 
                 \coordinate (c) at ($ (center2) - .1*(1,0) $);
                 \draw[red][line width=0.1mm] (c) -- (b); 
                 \coordinate (d) at ($ (center2) + .2*(0,1) $); 
			     \coordinate (center3) at (2.99,2.8); 
                 \coordinate (b) at ($ (center3) + .1*(1,0) $); 
                 \coordinate (c) at ($ (center3) - .1*(1,0) $);
                 \draw[red][line width=0.1mm] (c) -- (b); 
                 \draw[dashed][->,>=stealth,red][line width=0.1mm] (center3) -- (center2); 
                \draw[dashed][->,>=stealth,red][line width=0.1mm] (center2) -- (center3); 
                 \coordinate (a) at (0,0); 
                 \node[] at (a) {\tiny.};
                 \coordinate (center2) at (6.9,3.1); 
                 \coordinate (b) at ($ (center2) + 0.26*(1,0) $);
                 \coordinate (c) at ($ (b) + 0.36*(0,1) $);
                 \coordinate (d) at ($ (c) - 0.26*(1,0) $);
                 \draw[line width=0.2mm,dashdotted] (center2) -- (b); 
                 \draw[line width=0.2mm,dashdotted] (b) -- (c); 
                 \draw[line width=0.2mm,dashdotted] (d) -- (c); 
                 \draw[line width=0.2mm,dashdotted] (d) -- (center2); 
                 \coordinate (e) at ($ (d) - 2.3*(1,.1) $);
                 \draw[line width=0.2mm,dashdotted] (d) -- (e); 
                 \coordinate (a) at (0,0); 
                 \node[] at (a) {\tiny.};
                 \coordinate (center3) at (6.9,0.28); 
                 \coordinate (b) at ($ (center3) + 0.26*(1,0) $);
                 \coordinate (c) at ($ (b) + 1.57*(0,1) $);
                 \coordinate (d) at ($ (c) - 0.26*(1,0) $);
                 \draw[line width=0.2mm,red,dashdotted] (center3) -- (b); 
                 \draw[line width=0.2mm,red,dashdotted] (b) -- (c); 
                 \draw[line width=0.2mm,red,dashdotted] (d) -- (c); 
                 \draw[line width=0.2mm,red,dashdotted] (d) -- (center3); 
                 \coordinate (e1) at ($ (d) - 2.3*(1,.2) $);
                 \draw[line width=0.2mm,red,dashdotted] (d) -- (e1); 

             \end{tikzpicture}
		\end{overpic}
	\end{center}
	\caption{Results of stress-controlled bi-axial tests. Evolution of (\textbf{a}) the strain $\epsilon$ and (\textbf{b}) the tangent compliance $\dot\epsilon$ with the imposed stress $\sigma$ prior to the failure point at $p_0=0.01$, $\mu=0.1$, and $N=80\times 80$. The insets are the close-up views of the main graphs. The dashed line indicates the noise floor $\dot{\epsilon}_c$. The avalanche initiation, duration, and its magnitude are indicated by $\sigma_i$, $\Delta\sigma$, and $S$, respectively. }
	\label{fig:stressCtrl}
\end{figure}
Prior to shearing, samples were prepared by assigning $N$ particles randomly in a bi-periodic $L\times L$ square box which was then pre-compressed isotropically using a Berendsen barostat to a target stress $\sigma_{yy}=\sigma_{xx}=-p_0$.
Here the bulk stress tensor 
\begin{equation}
\sigma_{\alpha\beta}=L^{-d}\sum_i\sum_{i<j}(\vec{f}_{ij}\otimes\vec{r}_{ij}) _{\alpha\beta},
\end{equation}
is defined using the Kirkwood-Irvine expression \cite{allen2017computer} where $\vec{f}_{ij}=\vec{f}_n+\vec{f}_t$ and $\vec{r}_{ij}=\vec{r}_i-\vec{r}_j$.
Please note that grains were assumed frictionless during compaction.
A strain-controlled condition was subsequently applied  by deforming the periodic box along $y$ at a constant axial strain rate $\dot\epsilon_{yy}$. 
During the loading phase, the simulation box remained coupled to the barostat along $x$ with $\sigma_{xx}$ retained at the pre-compression level $-p_0$.
The numerical range of the applied pressure $p_0$ we have tested corresponds to the average overlap $\frac{\delta}{R_s}\simeq0.01-0.1$.
We also checked that $\dot\epsilon_{yy}$ was slow enough that the stress condition was almost insensitive to the loading rate.

A number of stress-controlled tests was also carried out where barostating was applied on both dimensions $x$ and $y$ with $\sigma_{xx}=-p_0$ and $\sigma_{yy}=-p_0-2\sigma$. 
Here $2\sigma$ represents the normal stress difference being applied at a slow rate ${\dot\sigma}$.
Upon quasi-static loading, the shear response $\epsilon$ shows remarkable softening in Fig.~\ref{fig:stressCtrl}(a) that evolves smoothly at low stress $\sigma$ but then reveals intermittent features as the deformation proceeds.
The fluctuations are more pronounced in the (tangent) compliance $\dot{\epsilon}\doteq\frac{\partial\epsilon}{\partial\sigma}$ as in Fig.~\ref{fig:stressCtrl}(b) with a noise floor $\dot{\epsilon}_{c}$ that is frequently interrupted by narrowly distributed peaks termed as \emph{avalanches}.
The avalanche size $S=\int_{\sigma_i}^{\sigma_i+\Delta\sigma}\sigma\dot{\epsilon}~d\sigma$ has dimensions of energy per unit volume and corresponds to an avalanche initiating at $\sigma_i$ with duration $\Delta\sigma$.
We quantified the temporal fluctuations of $S$ in terms of the statistical mean $\langle S\rangle$ and maximum size $S_\text{max}$ computed over stress bins that are typically one order of magnitude larger than the discretization value $\dot{\sigma}\Delta t$.
%
\section{Damage-based Analysis}\label{sec:DamageAnalysis}
The analysis was made by perturbing the stressed samples via a low amplitude-low frequency shear $\delta\sigma~\text{sin}(\omega_{\sigma}t)$ displayed in Fig.~\ref{fig:Loadunload}(a).
Here the magnitude of perturbation is $\delta\sigma\simeq \dot{\sigma}\Delta t$ with the frequency of oscillations $\omega_{\sigma}\simeq0.1~\omega_n$.
Upon periodic loading in Fig.~\ref{fig:Loadunload}(b), the shear response finally enters a limit cycle that enables for measurements of the associated shear stiffness $G$.
%
\begin{figure}[t]
	\begin{center}
		\begin{overpic}[width=8.6cm]{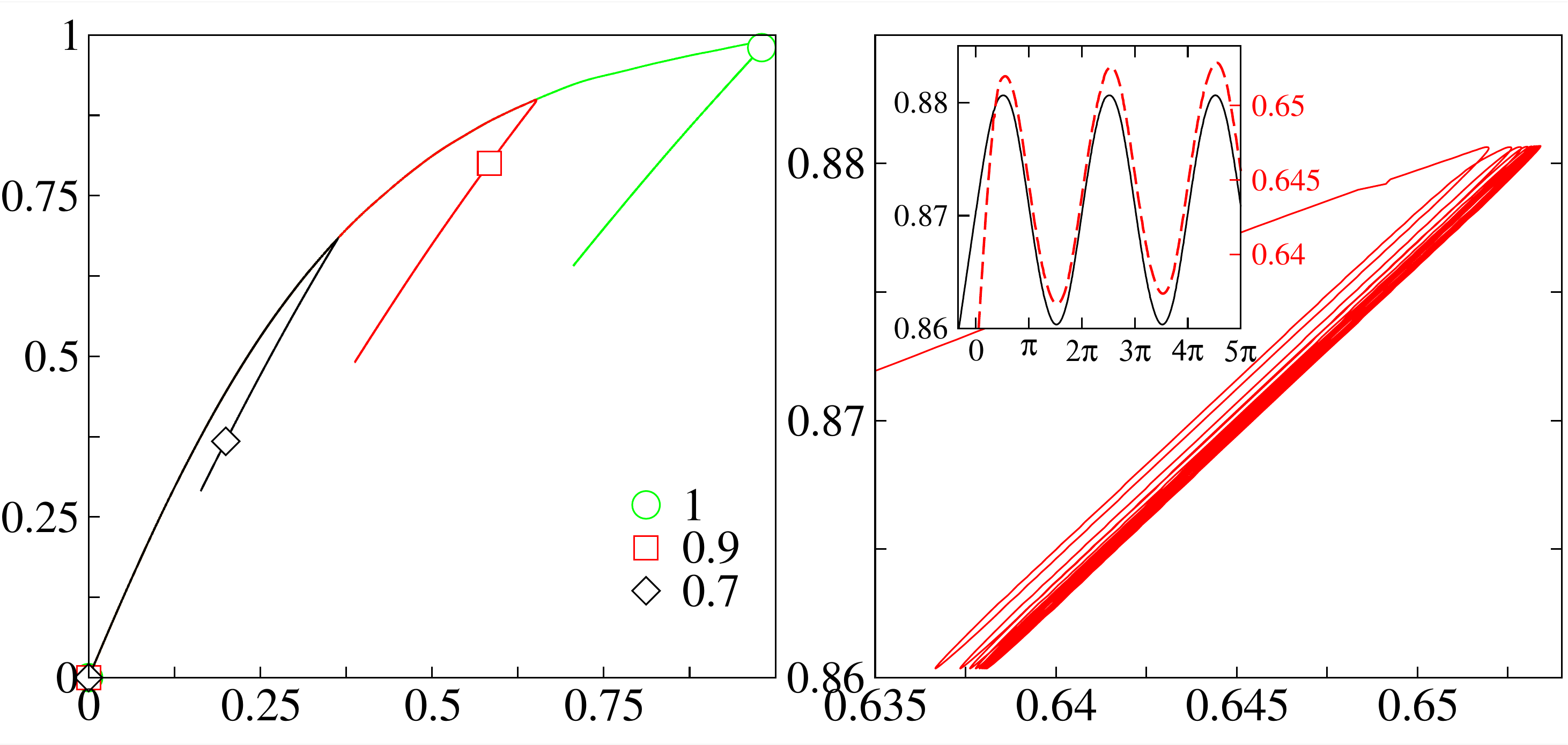}
            \put (27,-2) {$\bar\epsilon=\frac{\epsilon}{\epsilon_f}$} 
            \put (42,18) {\small$\bar\sigma$} 
            \put (68,21) {\small$\omega_\sigma t$} 
            \put (35,31) {\sffamily\setlength{\fboxsep}{0pt}\colorbox{white}{\strut\bfseries\textcolor{red}{\small\begin{turn}{90}{\small$\delta\sigma$}\end{turn}}}} 
            \put (62,37) {\sffamily\setlength{\fboxsep}{0pt}\colorbox{white}{\strut\bfseries\textcolor{black}{\small\begin{turn}{90}{\small$\bar\sigma$}\end{turn}}}} 
            \put (85,37) {\sffamily\setlength{\fboxsep}{0pt}\colorbox{white}{\strut\bfseries\textcolor{red}{\small\begin{turn}{90}{\small$\bar\epsilon$}\end{turn}}}} 
            \put (-4,20) {\sffamily\setlength{\fboxsep}{0pt}\colorbox{white}{\strut\bfseries\textcolor{black}{\small\begin{turn}{90}{$\bar\sigma=\frac{\sigma}{\sigma_f}$}\end{turn}}}} 
            \put (94,22) {\sffamily\setlength{\fboxsep}{0pt}\colorbox{white}{\strut\bfseries\textcolor{black}{\small\begin{turn}{90}{$\bar\sigma=\frac{\sigma}{\sigma_f}$}\end{turn}}}} 
            \put (74,-2) {$\bar\epsilon=\frac{\epsilon}{\epsilon_f}$} 
            \put (7,7) {\sffamily\setlength{\fboxsep}{0pt}\colorbox{black}{\strut\bfseries\textcolor{white}{\small$(a)$}}} 
            \put (57,7) {\sffamily\setlength{\fboxsep}{0pt}\colorbox{black}{\strut\bfseries\textcolor{white}{\small$(b)$}}} 
            \begin{tikzpicture}
                \coordinate (a) at (0,0); 
                \node[] at (a) {\tiny.};
                \coordinate (center1) at (6.6,1.4); 
                \coordinate (b) at ($ (center1) + 0.5*(1,1.0) $);
                \coordinate (c) at ($ (center1) + 0.5*(1,0) $); 
                \draw[red][line width=0.1mm] (center1) -- (b); 
                \draw[red][line width=0.1mm] (center1) -- (c); 
                \draw[red][line width=0.1mm] (b) -- (c); 
                \node[red] at ($(b)-(0.15,.6)$) {\tiny{1}}; 
                \node[red] at ($(b)-(-.2,0.25)$) {\tiny{$2G$}}; 
                \coordinate (center1) at (.92,2.5); 
                \coordinate (b) at ($ (center1) + 0.45*(.5,0) $);
                \coordinate (c) at ($ (center1) + 0.5*(0,-1) $); 
                \draw[line width=0.1mm] (center1) -- (b); 
                \draw[line width=0.1mm] (center1) -- (c); 
                \draw[line width=0.1mm] (b) -- (c); 
                \node at ($(b)-(0.13,-0.1)$) {\tiny{1}}; 
                \node at ($(b)-(0.45,0.2)$) {\tiny{$2G_0$}}; 
           	   \coordinate (center1) at (0.33,0.25); 
                \coordinate (b) at ($ (center1) + 1.2*(1,2.5) $);
                \draw[dashdotted][line width=0.1mm] (center1) -- (b); 
			     \coordinate (center2) at (2.8,3.43); 
                \coordinate (b) at ($ (center2) + .2*(1,0) $); 
                \coordinate (c) at ($ (center2) - .2*(1,0) $);
                \draw[red][line width=0.1mm] (c) -- (b); 
                \coordinate (d) at ($ (center2) + .2*(0,1) $); 
                \draw[->,>=stealth,red][line width=0.1mm] (d) -- (center2); 
			     \coordinate (center3) at (2.8,2.0); 
                \coordinate (b) at ($ (center3) + .2*(1,0) $); 
                \coordinate (c) at ($ (center3) - .2*(1,0) $);
                \draw[red][line width=0.1mm] (c) -- (b); 
                \draw[dashed][->,>=stealth,red][line width=0.1mm] (center3) -- (center2); 
                \draw[dashed][->,>=stealth,red][line width=0.1mm] (center2) -- (center3); 
			     \coordinate (center1) at (2.24,2.65); 
                \coordinate (b) at ($ (center1) - 0.1*(1,1.6) $); 
                \draw[->,>=stealth,red][line width=0.1mm] (center1) -- (b); 
                \coordinate (b) at ($ (center1) + 0.1*(1,1.6) $); 
                \draw[->,>=stealth,red][line width=0.1mm] (center1) -- (b); 
			     \coordinate (center1) at (1.42,2.31); 
                \coordinate (b) at ($ (center1) + 0.001*(1,1.5) $); 
                \draw[->,>=stealth,black][line width=0.1mm] (center1) -- (b); 
            \end{tikzpicture}
		\end{overpic}
	\end{center}
	\caption{Stress-controlled oscillatory shear perturbation. (\textbf{a}) The stress paths in terms of the normalized stress $\bar\sigma$ and shear strain $\bar\epsilon$ prior to the failure point $(\epsilon_f,\sigma_f)$ at three different stress levels. The arrows indicate the maximum stress and loading directions in the graph. The amplitude of oscillations is denoted by $\delta\sigma$. The slope of the dashed line measures the shear modulus of the intact state $G_0$. (\textbf{b}) Close-up view of the loading-unloading sequence and the associated modulus $G$. The inset shows the phase evolution of $\bar\sigma$ and $\bar\epsilon$ with $\omega_{\sigma}$ being the frequency of shear oscillations.}
	\label{fig:Loadunload}
\end{figure}
\begin{figure}[t]
	\begin{center} 
		\begin{overpic}[width=8.6cm]{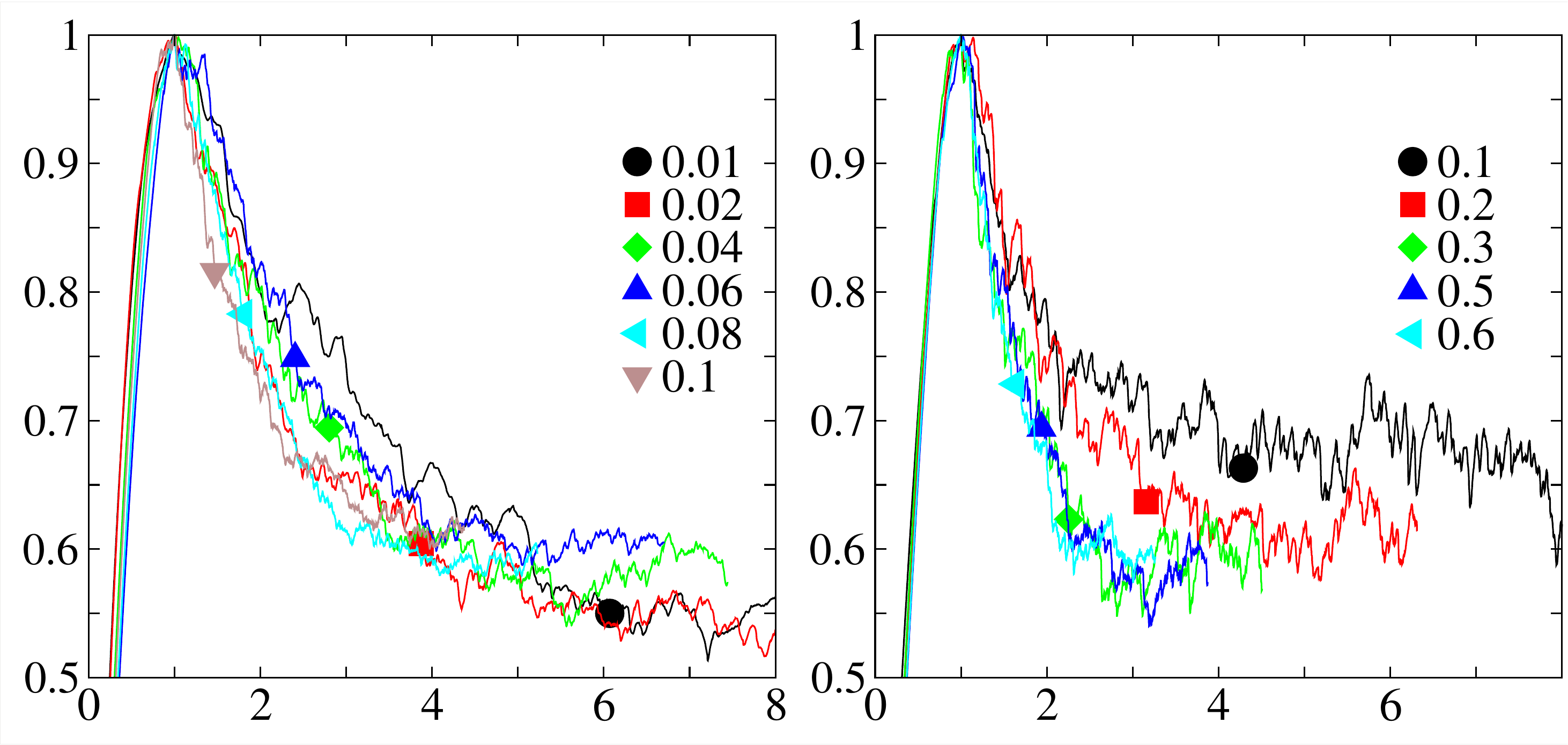}
              \put (91,40) {$\mu$} 
              \put (42,40) {$p_0$} 
              \put (27,-2) {$\bar\epsilon=\frac{\epsilon}{\epsilon_f}$} 
              \put (76,-2) {$\bar\epsilon=\frac{\epsilon}{\epsilon_f}$} 
              \put (-4,20) {\sffamily\setlength{\fboxsep}{0pt}\colorbox{white}{\strut\bfseries\textcolor{black}{\small\begin{turn}{90}{$\bar\sigma=\frac{\sigma}{\sigma_f}$}\end{turn}}}} 
              \put (58,20) {\sffamily\setlength{\fboxsep}{0pt}\colorbox{white}{\strut\bfseries\textcolor{black}{\small\begin{turn}{90}{$\bar\sigma=\frac{\sigma}{\sigma_f}$}\end{turn}}}} 
             \put (20,40) {\sffamily\setlength{\fboxsep}{0pt}\colorbox{black}{\strut\bfseries\textcolor{white}{\small~$\mu=0.4$~}}} 
             \put (69,40) {\sffamily\setlength{\fboxsep}{0pt}\colorbox{black}{\strut\bfseries\textcolor{white}{\small~$p_0=0.04$~}}} 
             \put (7,7) {\sffamily\setlength{\fboxsep}{0pt}\colorbox{black}{\strut\bfseries\textcolor{white}{\small$(a)$}}} 
             \put (57,7) {\sffamily\setlength{\fboxsep}{0pt}\colorbox{black}{\strut\bfseries\textcolor{white}{\small$(b)$}}} 
             \begin{tikzpicture}
                 \coordinate (a) at (0,0); 
                 \node[] at (a) {\tiny.};
 			     \coordinate (center1) at (1.9,2); 
                  \coordinate (b) at ($ (center1) - .6*(1,1.4) $); 
                 \draw[->,>=stealth,black][line width=0.2mm] (center1) -- (b); 
               		\coordinate (center1) at (6.1,2); 
                  \coordinate (b) at ($ (center1) - .6*(1,1.4) $); 
                 \draw[->,>=stealth,black][line width=0.2mm] (center1) -- (b); 
             \end{tikzpicture}

		\end{overpic}
	\end{center}
	\caption{Evolution of the bulk (normalized) stress $\bar{\sigma}$ as a function of (normalized) strain $\bar\epsilon$ with (\textbf{a}) $\mu=0.4$ at multiple $p_0$ and (\textbf{b}) $p_0=0.04$ at several friction coefficients with the system size $N=80\times80$. Here $\sigma_f$ and $\epsilon_f$ denote the peak stress and the corresponding strain, respectively. The control parameter is the axial strain $\epsilon_{yy}$. The arrows indicate the increasing pressure in (a) and friction value in (b).}
	\label{fig:sigmaEsplnAppndx}
\end{figure}
\begin{figure}[t]
	\begin{center} 
		\begin{overpic}[width=8.6cm]{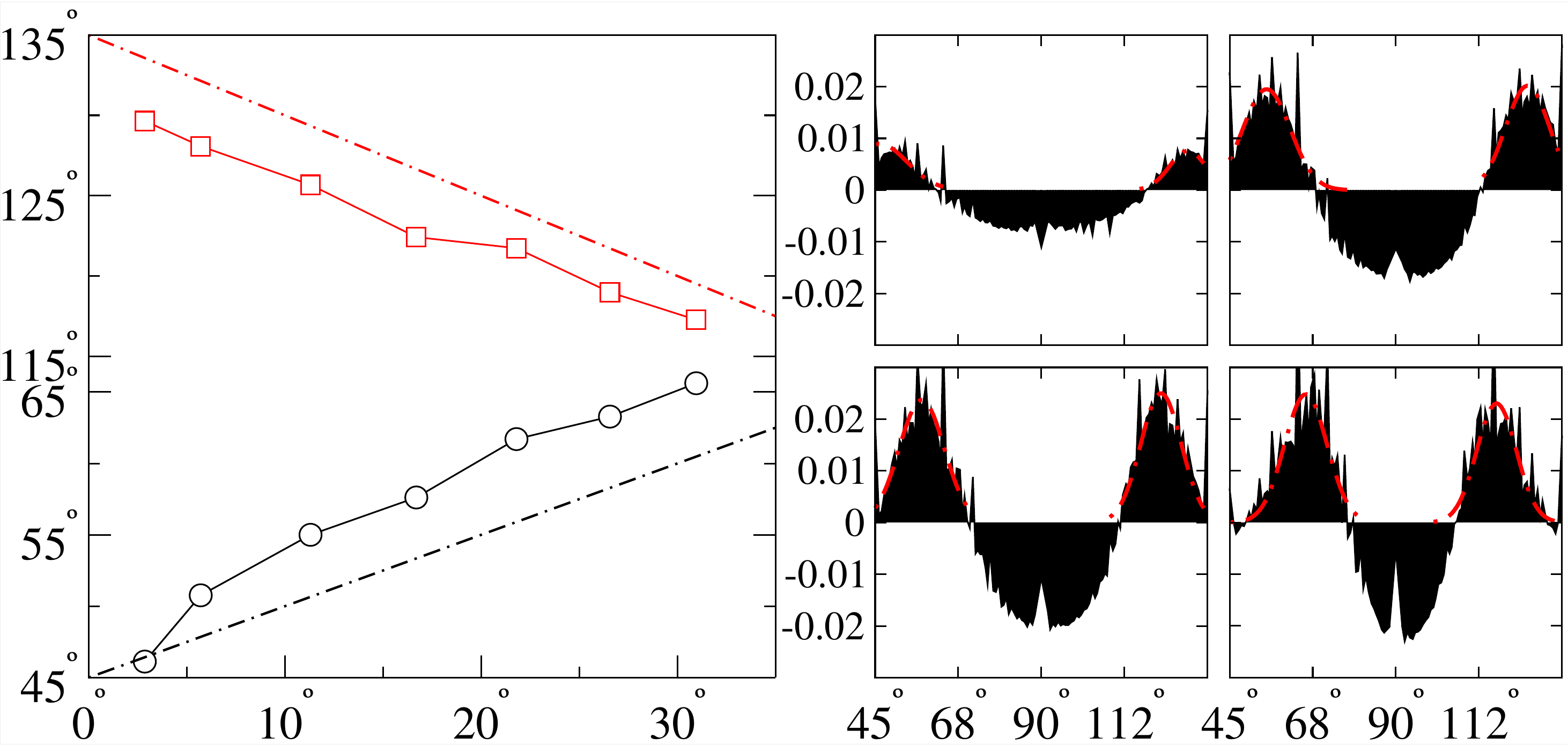} 
              \put (26,-2) {$\phi$} 
              \put (65,-3) {$\theta$} 
              \put (88,-3) {$\theta$} 
              \put (-4,23) {\sffamily\setlength{\fboxsep}{0pt}\colorbox{white}{\strut\bfseries\textcolor{black}{\small\begin{turn}{90}{$\theta_\text{peak}$}\end{turn}}}} 
              \put (101,10.5) {\sffamily\setlength{\fboxsep}{0pt}\colorbox{white}{\strut\bfseries\textcolor{black}{\small\begin{turn}{90}{$C_\epsilon(\theta)$}\end{turn}}}} 
              \put (101,30.5) {\sffamily\setlength{\fboxsep}{0pt}\colorbox{white}{\strut\bfseries\textcolor{black}{\small\begin{turn}{90}{$C_\epsilon(\theta)$}\end{turn}}}} 
             \put (22,7) {\small\begin{turn}{20}{$45^\circ+\frac{\phi}{2}$~}\end{turn}} 
             \put (22,40.5) {\color{red}{\small\begin{turn}{-20}{$135^\circ-\frac{\phi}{2}$~}\end{turn}}} 
             \put (43.5,41.5) {\sffamily\setlength{\fboxsep}{0pt}\colorbox{black}{\strut\bfseries\textcolor{white}{\small$(a)$}}} 
             \put (57,28) {\sffamily\setlength{\fboxsep}{0pt}\colorbox{black}{\strut\bfseries\textcolor{white}{\small$(b)$}}} 
             \put (79.5,28) {\sffamily\setlength{\fboxsep}{0pt}\colorbox{black}{\strut\bfseries\textcolor{white}{\small$(c)$}}} 
             \put (79.5,7) {\sffamily\setlength{\fboxsep}{0pt}\colorbox{black}{\strut\bfseries\textcolor{white}{\small$(e)$}}} 
             \put (57,7) {\sffamily\setlength{\fboxsep}{0pt}\colorbox{black}{\strut\bfseries\textcolor{white}{\small$(d)$}}} 
          \end{overpic}
	\end{center}
		\begin{overpic}[width=8.6cm]{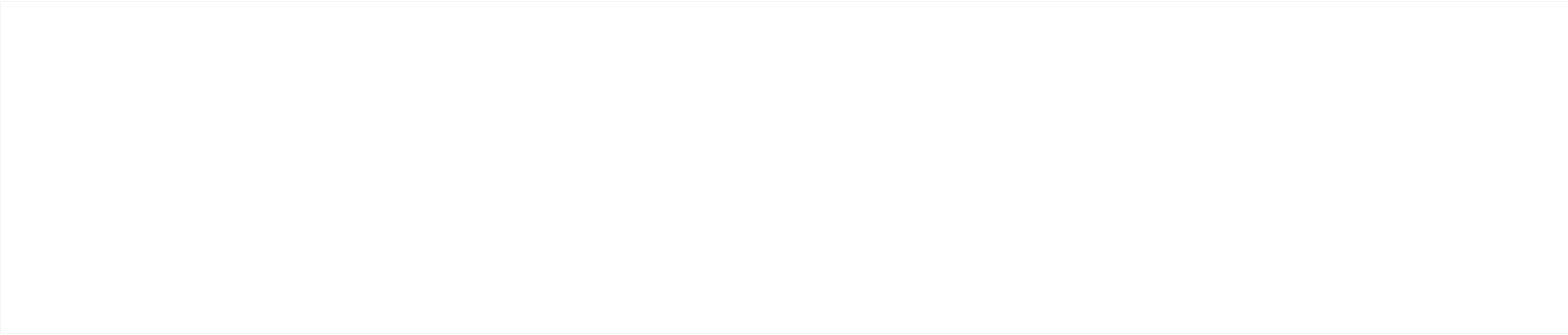} 
		 		    \put(8.25,.8){\includegraphics[height=1.7cm,width=1.8cm,frame]{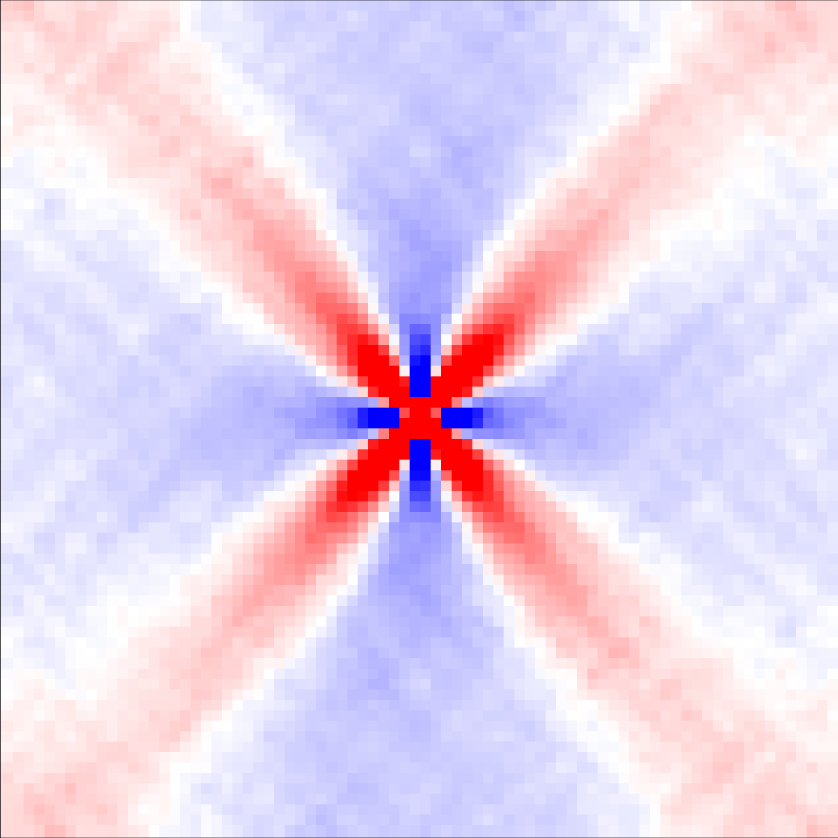}}
		 		    \put(30.8,.8){\includegraphics[height=1.7cm,width=1.8cm,frame]{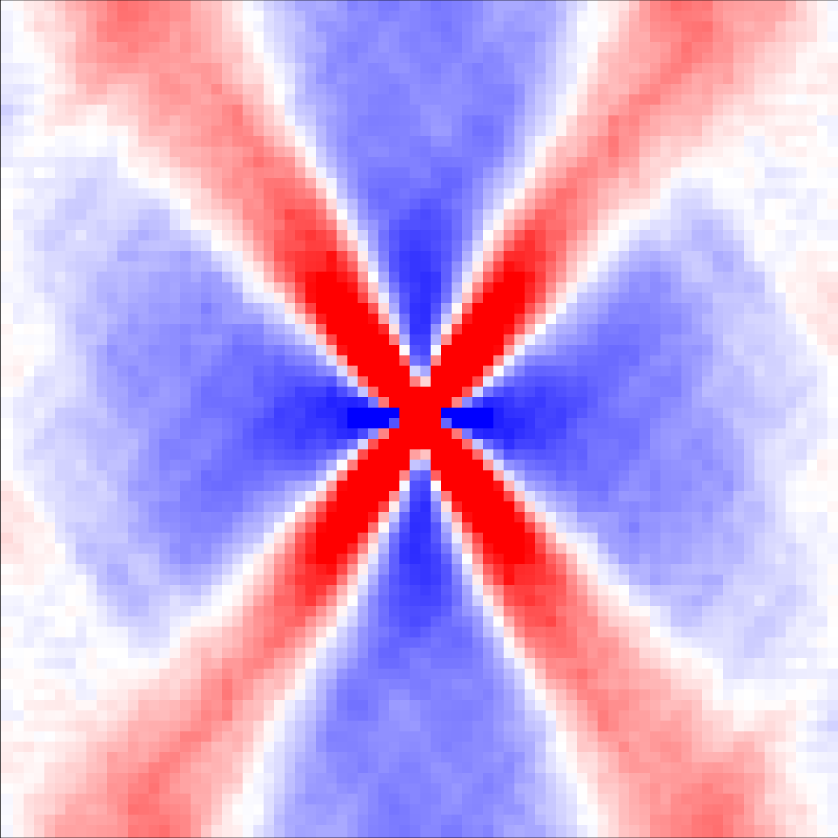}}
		 		    \put(53.35,.8){\includegraphics[height=1.7cm,width=1.8cm,frame]{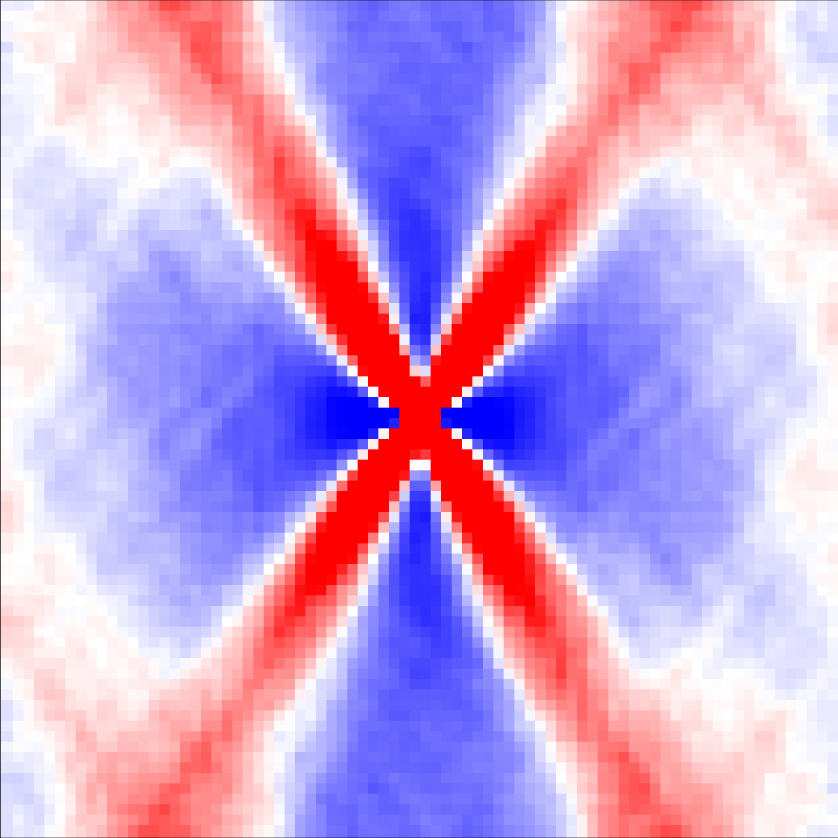}}
		 		    \put(75.9,.8){\includegraphics[height=1.7cm,width=1.8cm,frame]{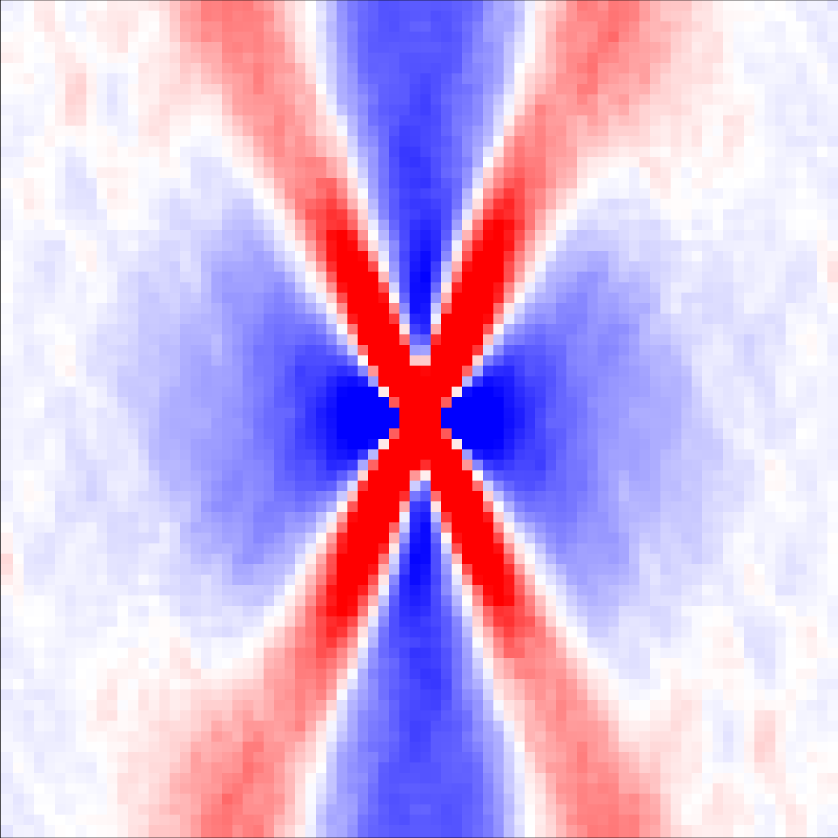}}
             \put (9,17) {\sffamily\setlength{\fboxsep}{0pt}\colorbox{black}{\strut\bfseries\textcolor{white}{\small$(f)$}}} 
             \put (31.55,17) {\sffamily\setlength{\fboxsep}{0pt}\colorbox{black}{\strut\bfseries\textcolor{white}{\small$(g)$}}} 
             \put (54.1,17) {\sffamily\setlength{\fboxsep}{0pt}\colorbox{black}{\strut\bfseries\textcolor{white}{\small$(h)$}}} 
             \put (76.65,17) {\sffamily\setlength{\fboxsep}{0pt}\colorbox{black}{\strut\bfseries\textcolor{white}{\small$(i)$}}} 
 		    \put(42.75,-3){\includegraphics[height=0.15cm,width=1.7cm]{./colorBar.png}}
			\put (38,-3) {\color{black}\tiny$-1$} 
			\put (63.5,-3) {\color{black}\tiny$1(\times10^{-2})$} 
			\begin{tikzpicture} 
				\coordinate (a) at (0,0); 
				\node[] at (a) {\tiny.}; 
				\coordinate (center) at (1.48,.8); 
				\draw[dashdotted,thick] (center) -- ( $ (center) + .5*(1,1.0512) $); 
				\draw[dashdotted,thick] (center) -- ( $ (center) + .5*(1,0) $); 
				\coordinate (center) at (3.42,.8); 
				\draw[dashdotted,thick] (center) -- ( $ (center) + .5*(1,1.2198) $); 
				\draw[dashdotted,thick] (center) -- ( $ (center) + .5*(1,0) $); 
				\coordinate (center) at (5.36,.8); 
				\draw[dashdotted,thick] (center) -- ( $ (center) + .5*(1,1.344) $); 
				\draw[dashdotted,thick] (center) -- ( $ (center) + .5*(1,0) $); 
				\coordinate (center) at (7.3,.8); 
				\draw[dashdotted,thick] (center) -- ( $ (center) + .5*(1,1.766) $); 
				\draw[dashdotted,thick] (center) -- ( $ (center) + .5*(1,0) $); 
				\coordinate (center2) at (7.6,0.15); 
				\draw[thick] (center2) -- ( $ (center2) + 0.5*(1,0) $); 
				\draw[thick] ($ (center2) + 0.5*(1,0)-0.1*(0,1) $) -- ( $ (center2) + 0.5*(1,0)+0.1*(0,1) $); 
				\draw[thick] ($ (center2) -0.1*(0,1) $) -- ( $ (center2) +0.1*(0,1) $); 
				\node[] at ($(center2)+0.5*(1,0)+(-0.25,0.2)$) {\tiny{$50R_s$}}; 

			\end{tikzpicture} 
          \end{overpic}
	\caption{Anisotropic part of the strain correlation function $C_\epsilon(\theta)$ and its dependence on the friction angle $\phi=\text{tan}^{-1}\mu$. (\textbf{a}) Locations of the peaks in $C_\epsilon(\theta)$ denoted by $\theta_\text{peak}$ versus $\phi$. The dash-dotted lines in the plot designate Mohr-Coulomb predictions. (\textbf{b}-\textbf{e}) $C_\epsilon(\theta)$ plotted against $\theta$ at $\phi\simeq 3^\circ, 11^\circ, 17^\circ, 31^\circ$ corresponding to $\mu=0.05,0.2,0.3,0.6$. Gaussian fits are denoted by the dashed curves. (\textbf{f}-\textbf{i}) Strain correlation maps. The dash-dotted lines in the strain maps are Mohr-Coulomb orientations $45^\circ+\frac{\phi}{2}$. The correlation analysis is similar to the one presented in Fig.~4 of the main text.} 
	\label{fig:MohrCoulombAppndx}
\end{figure}

\section{Stress discontinuity}\label{sec:StressDiscontinuity} The results of the compression test plotted in Fig.~\ref{fig:sigmaEsplnAppndx}(a) at $\mu=0.4$ under different imposed pressures $p_0$ and also multiple friction levels at constant pressure $p_0=0.04$ of Fig.~\ref{fig:sigmaEsplnAppndx}(b) demonstrate that the nature of bulk response will depend on the two control parameters $p_0$ and $\mu$.  
In both graphs, the stress overshoot tends to become more prominent in strongly frictional and/or confined aggregates with the former being even more effective than the latter as suggested in Fig.~\ref{fig:sigmaEsplnAppndx}(b).
Indeed, similar numerical observations were made within the context of glass rheology and its dependence on preparation protocols \cite{ozawa2018random}. 
More specifically, the yielding regime was found to be sensitive to the degree of \emph{annealing} in prepared samples which itself regulates heterogeneities in the glass structure. 
Based on this picture, the emergence of shear localization in ``well-aged'' glasses, accompanying the macroscopic stress drop, was interpreted in the framework of the non-equilibrium phase coexistence and first order discontinuous transition.
These findings were independently validated in the context of the elasto-plastic models and mean-field estimations in \cite{popovic2018elasto} by arguing that the statistical distribution of local instability thresholds will depend on the extent of structural heterogeneities in the quenched glass and, therefore, is accountable for the brittle fracture-like transition.

\section{On the validity of the Mohr-Coulomb framework}\label{sec:MohrCoulomb} Figure~\ref{fig:MohrCoulombAppndx} displays the anisotropic part of the correlation function $C_\epsilon(\theta)$, averaged $C_{\epsilon}(\vec{r}-\vec{r}^{\hspace{1pt}\prime})$ over distances $|\vec{r}-\vec{r}^{\hspace{1pt}\prime}|$, at multiple friction angles $\phi=\tan^{-1}\mu$. 
There are two marked maxima in each data set of Fig.~\ref{fig:MohrCoulombAppndx}(b-e) that correspond to the positive sectors of the correlation maps illustrated in Fig.~\ref{fig:MohrCoulombAppndx}(f-i).
We quantify the positions of these peaks by fitting a sum of two Gaussian peaks to the data which are denoted by $\theta_\text{peak}$ in Fig.~\ref{fig:MohrCoulombAppndx}(a).
We find that the theory does not apply because slip lines will not occur on planes with critical ratio between the resolved shear and normal stress as stated by the Mohr-Coulomb criterion.
\end{document}